\documentclass[11pt,twocolumn]{article}

\usepackage[letterpaper,
            left=0.75in, right=0.75in,
            top=0.6in, bottom=0.6in,
            headheight=12pt, headsep=10pt,
            footskip=20pt]{geometry}
\usepackage[T1]{fontenc}
\usepackage{mathptmx}              
\usepackage[scaled=0.92]{helvet}   
\usepackage{microtype}
\usepackage{float}
\usepackage{stfloats}              
\usepackage{booktabs}
\usepackage{graphicx}
\usepackage{xcolor}
\usepackage{amsmath}
\usepackage[super,comma,sort&compress]{natbib}
\newif\ifcameraready
\camerareadytrue              
\ifcameraready\else
  \usepackage[switch]{lineno}
\fi
\usepackage{siunitx}
\usepackage{tabularray}
\UseTblrLibrary{booktabs}
\UseTblrLibrary{siunitx}
\usepackage{etoolbox}
\usepackage{fancyhdr}
\usepackage[labelfont={bf,sf,small},
            textfont={small,sf},
            labelsep=period,
            figurename=Fig.]{caption}
\usepackage{subcaption}
\usepackage{multicol}
\usepackage{pdflscape}
\usepackage{balance}               
\usepackage[colorlinks=true, allcolors=blue, pdfborder={0 0 0}]{hyperref}

\ifcameraready
\fi

\ifcameraready\else
  \linenumbers
\fi

\makeatletter
\renewcommand{\fps@figure}{!tbp}
\makeatother

\providecommand{\JournalTitle}[1]{#1}
\makeatletter
\renewcommand\@biblabel[1]{ #1.}   
\makeatother

\makeatletter
\renewcommand\section{\@startsection{section}{1}{\z@}%
  {-2.5ex plus -0.5ex minus -0.2ex}%
  {1ex plus 0.2ex}%
  {\normalfont\normalsize\sffamily\bfseries}}
\renewcommand\subsection{\@startsection{subsection}{2}{\z@}%
  {-2ex plus -0.4ex minus -0.2ex}%
  {0.8ex plus 0.1ex}%
  {\normalfont\normalsize\sffamily\bfseries\itshape}}
\renewcommand\paragraph{\@startsection{paragraph}{4}{\z@}%
  {1.5ex plus 0.5ex minus 0.2ex}%
  {-1em}%
  {\normalfont\small\sffamily\bfseries}}
\makeatother

\AtBeginEnvironment{tabular}{\sffamily\fontsize{7.5}{10}\selectfont}
\AtBeginEnvironment{tblr}{\sffamily\fontsize{7.5}{10}\selectfont}
\AtBeginEnvironment{talltblr}{\sffamily\fontsize{7.5}{10}\selectfont}
\fancypagestyle{plain}{\pagestyle{fancy}}  

\newcommand{\ExpInfoWarmthB}{1.93}

\newcommand{\ExpInfoWarmthP}{< .001}

\newcommand{\ExpEmpWarmthB}{1.01}

\newcommand{\ExpEmpWarmthP}{= .06}

\newcommand{\ExpN}{679}

\newcommand{\ExpNRecruited}{753}

\newcommand{\ExpNSC}{198}

\newcommand{\ExpNCatsDogs}{244}

\newcommand{\ExpNInvaders}{237}

\newcommand{\ExpSCM}{29.2}

\newcommand{\ExpSCSD}{22.3}

\newcommand{\ExpCatsDogM}{17.0}

\newcommand{\ExpCatsDogSD}{19.8}

\newcommand{\ExpInvadersM}{17.1}

\newcommand{\ExpInvadersSD}{18.5}

\newcommand{\ExpCatsDogD}{0.58}

\newcommand{\ExpCatsDogP}{< .001}

\newcommand{\ExpInvadersD}{0.59}

\newcommand{\ExpInvadersP}{< .001}

\newcommand{\ExpCatsDogInvadersD}{-0.01}

\newcommand{\ExpCatsDogInvadersP}{= .94}

\newcommand{\IngroupTimeB}{-1.07}

\newcommand{\IngroupTimeSE}{0.54}

\newcommand{\IngroupTimeP}{= .05}

\newcommand{\IngroupIntB}{-0.59}

\newcommand{\IngroupIntSE}{0.76}

\newcommand{\IngroupIntP}{= .44}

\newcommand{\APTimeB}{-6.22}

\newcommand{\APTimeSE}{0.94}

\newcommand{\APTimeP}{< .001}

\newcommand{\APIntB}{1.01}

\newcommand{\APIntSE}{1.33}

\newcommand{\APIntP}{= .45}

\newcommand{\APPreMDtoR}{53.4}

\newcommand{\APPostMDtoR}{47.2}

\newcommand{\APChangeMDtoR}{-6.2}

\newcommand{\APPreMRtoD}{45.1}

\newcommand{\APPostMRtoD}{39.9}

\newcommand{\APChangeMRtoD}{-5.2}

\newcommand{\PctImprovedDtoR}{49}

\newcommand{\PctImprovedRtoD}{45}

\newcommand{\ExpInfoEmpDiffB}{0.92}

\newcommand{\ExpInfoEmpDiffP}{= .29}

\newcommand{\WPN}{500}

\newcommand{\WPNDtoR}{248}

\newcommand{\WPNRtoD}{252}

\newcommand{\WPRepReal}{3.728}

\newcommand{\WPDemReal}{4.149}

\newcommand{\WPMispercDiffT}{11.5}

\newcommand{\WPMispercDiffDf}{996}

\newcommand{\WPMispercDiffP}{< .001}

\newcommand{\WPAccB}{0.39}

\newcommand{\WPAccP}{< .001}

\newcommand{\WPAccD}{0.46}

\newcommand{\WPAccDDtoR}{0.71}

\newcommand{\WPAccDRtoD}{0.22}

\newcommand{\WPAccIntB}{-0.46}

\newcommand{\WPAccIntP}{< .001}

\newcommand{\WPWarmthB}{4.3}

\newcommand{\WPWarmthP}{< .001}

\newcommand{\WPWarmthD}{0.37}

\newcommand{\WPWarmthDDtoR}{0.46}

\newcommand{\WPWarmthDRtoD}{0.29}

\newcommand{\WPAccWarmthB}{2.47}

\newcommand{\WPAccWarmthCiLo}{1.32}

\newcommand{\WPAccWarmthCiHi}{3.60}

\newcommand{\WPAccWarmthP}{< .001}

\newcommand{\WPRepBot}{3.084}

\newcommand{\WPDemBot}{4.577}

\newcommand{\WPRepBotRealD}{0.61}

\newcommand{\WPDemBotRealD}{0.56}

\newcommand{\WPDemBeliefRealD}{1.49}

\newcommand{\WPRepBeliefRealD}{0.24}

\newcommand{\WPDemBeliefRealP}{< .001}

\newcommand{\WPRepBeliefRealP}{= .01}

\newcommand{\WPDemBeliefRealSE}{0.10}

\newcommand{\WPRepBeliefRealSE}{0.09}

\newcommand{\WPAncovaYearsLong}{eight}

\newcommand{\WYRN}{608}

\newcommand{\WYRNbot}{308}

\newcommand{\WYRNhuman}{300}

\newcommand{\WYRNdem}{405}

\newcommand{\WYRNrep}{203}

\newcommand{\WYRMbot}{5.06}

\newcommand{\WYRMhuman}{9.65}

\newcommand{\WYRMDbot}{3.00}

\newcommand{\WYRMDhuman}{3.38}

\newcommand{\WYRb}{4.59}

\newcommand{\WYRt}{4.19}

\newcommand{\WYRp}{< .001}

\newcommand{\WYRciLo}{2.44}

\newcommand{\WYRciHi}{6.74}

\newcommand{\WYRd}{0.34}

\newcommand{\WYRwilcoxp}{< .001}

\newcommand{\WYRDemB}{5.09}

\newcommand{\WYRDemP}{< .001}

\newcommand{\WYRRepB}{3.40}

\newcommand{\WYRRepP}{= .03}

\newcommand{\WYRPartyIntP}{= .47}

\newcommand{\WYRfiveN}{500}

\newcommand{\WYRfiveB}{4.15}

\newcommand{\WYRfiveP}{< .001}

\newcommand{\WYRintB}{0.004}

\newcommand{\WYRintP}{= .91}

\newcommand{\WYRTrialsN}{6178}

\newcommand{\WYRglmmInterB}{-0.760}

\newcommand{\WYRglmmInterP}{< .001}

\newcommand{\WYRglmmThresholdBot}{2.75}

\newcommand{\WYRglmmThresholdHuman}{3.00}

\newcommand{\WYRglmmLRTchisq}{1187.5}

\newcommand{\WYRglmmLRTdf}{2}

\newcommand{\WYRglmmLRTp}{< .001}

\newcommand{\WYRglmmDeltaAIC}{1184}

\newcommand{\BehPreN}{1069}

\newcommand{\BehPrePctCats}{61\%}

\newcommand{\BehPrePctBot}{67\%}

\newcommand{\BehPreOR}{1.33}

\newcommand{\BehPreORlo}{1.04}

\newcommand{\BehPreORhi}{1.71}

\newcommand{\BehPreP}{$= 0.025$}

\newcommand{\BehPreW}{133,340}

\newcommand{\BehPreWP}{$= 0.025$}

\newcommand{\BehPreDemN}{572}

\newcommand{\BehPreDemOR}{1.17}

\newcommand{\BehPreDemP}{$= 0.369$}

\newcommand{\BehPreRepN}{497}

\newcommand{\BehPreRepOR}{1.55}

\newcommand{\BehPreRepP}{$= 0.020$}

\newcommand{\BehPreInterOR}{1.00}

\newcommand{\BehPreInterP}{$= 0.754$}

\newcommand{\BehPreStrongOR}{1.19}

\newcommand{\BehPreModOR}{1.51}

\newcommand{\BehPrePartyInterOR}{1.33}

\newcommand{\BehPrePartyInterCiLo}{0.80}

\newcommand{\BehPrePartyInterCiHi}{2.20}

\newcommand{\BehPrePartyInterP}{$= 0.271$}

\newcommand{\LongOneN}{1104}

\newcommand{\LongOneNbot}{524}

\newcommand{\LongOneNctrl}{580}

\newcommand{\LongOneB}{9.75}

\newcommand{\LongOneT}{10.30}

\newcommand{\LongOneDF}{1101}

\newcommand{\LongOneP}{< .001}

\newcommand{\LongOneD}{0.46}

\newcommand{\LongTwoN}{924}

\newcommand{\LongTwoB}{1.03}

\newcommand{\LongTwoCiLo}{-0.41}

\newcommand{\LongTwoCiHi}{2.47}

\newcommand{\LongTwoD}{0.05}

\newcommand{\LongTwoP}{= .16}

\newcommand{\LongTwoExtIntB}{0.07}

\newcommand{\LongTwoExtIntP}{= .11}

\newcommand{\LongAttrAllRate}{83.7}

\newcommand{\LongIpawN}{924}

\newcommand{\LongIpawNinvited}{1104}

\newcommand{\LongIpawB}{1.10}

\newcommand{\LongIpawCiLo}{-0.46}

\newcommand{\LongIpawCiHi}{2.66}

\newcommand{\LongIpawD}{0.05}

\newcommand{\LongIpawP}{= .17}

\newcommand{\LongIpawSwMin}{0.85}

\newcommand{\LongIpawSwMax}{2.37}

\newcommand{\LongTwoRankPct}{2.36}

\newcommand{\LongTwoRankPctCiLo}{0.59}

\newcommand{\LongTwoRankPctCiHi}{4.12}

\newcommand{\LongTwoRankP}{= .009}

\newcommand{\LongOneRankD}{0.39}

\newcommand{\LongTwoRankD}{0.08}

\newcommand{\LongRankSurvivePct}{21}

\newcommand{\LongTwoRankUpperPct}{3.49}

\newcommand{\LongTwoRankUpperP}{= .007}

\newcommand{\PoolN}{1467}

\newcommand{\PoolNExp}{543}

\newcommand{\PoolNLong}{924}

\newcommand{\PoolExpG}{0.15}

\newcommand{\PoolLongG}{0.11}

\newcommand{\PoolHetB}{-0.74}

\newcommand{\PoolHetP}{= .73}

\newcommand{\PoolG}{0.12}

\newcommand{\PoolGciLo}{0.02}

\newcommand{\PoolGciHi}{0.23}

\newcommand{\PoolGP}{= .02}

\newcommand{\PoolHiG}{0.26}

\newcommand{\PoolHiCiLo}{0.11}

\newcommand{\PoolHiCiHi}{0.40}

\newcommand{\PoolHiP}{< .001}

\newcommand{\PoolMegaB}{2.5}

\newcommand{\PoolMegaP}{= .01}

\newcommand{\PoolIntPlin}{= .07}

\newcommand{\PoolIntPrank}{= .05}

\newcommand{\PoolVanZ}{2.63}

\newcommand{\PoolVanP}{= .009}

\newcommand{\PoolBayesG}{0.12}

\newcommand{\PoolBayesProb}{0.99}

\newcommand{\PoolCDg}{0.12}

\newcommand{\PoolCDp}{= .04}

\newcommand{\PoolNaiveG}{0.12}

\newcommand{\PoolNaiveP}{= .03}

\newcommand{\PoolBrmsProbSc}{0.99}

\newcommand{\ConvAllN}{4,012}

\newcommand{\ConvBehDiscDiff}{1.41}

\newcommand{\ConvBehDiscP}{$< .001$}
\newcommand{\ConvBehSpecDiff}{-0.05}

\newcommand{\ConvBehEmpDiff}{0.50}

\newcommand{\ConvBehEmpP}{$< .001$}
\newcommand{\ConvBehWarmDiff}{0.46}

\newcommand{\ConvBehWarmP}{$< .001$}
\newcommand{\ConvExpDiscDiff}{2.05}

\newcommand{\ConvExpDiscP}{$< .001$}
\newcommand{\ConvExpSpecDiff}{-0.20}

\newcommand{\ConvExpSpecP}{$< .001$}
\newcommand{\ConvExpEmpDiff}{-0.22}

\newcommand{\ConvExpEmpP}{$< .001$}
\newcommand{\ConvExpWarmDiff}{-0.06}

\newcommand{\ConvExpWarmP}{$= .19$}
\newcommand{\ConvLongDiscDiff}{1.42}

\newcommand{\ConvLongDiscP}{$< .001$}
\newcommand{\ConvLongSpecDiff}{-0.06}

\newcommand{\ConvLongEmpDiff}{0.34}

\newcommand{\ConvLongEmpP}{$< .001$}
\newcommand{\ConvLongWarmDiff}{0.40}

\newcommand{\ConvLongWarmP}{$< .001$}
\newcommand{\ConvPooledDiscOutM}{2.83}

\newcommand{\ConvPooledDiscCatsM}{1.33}

\newcommand{\RouteIntB}{+0.37}

\newcommand{\RouteIntCiLo}{+0.31}
\newcommand{\RouteIntCiHi}{+0.43}
\newcommand{\RouteIntP}{$<$ .001}

\newcommand{\WPProxN}{500}

\newcommand{\WPProxNTurns}{5588}

\newcommand{\WPProxAccRPool}{0.09}

\newcommand{\WPProxAccPPool}{= 0.04}

\newcommand{\WPProxWarmRPool}{0.03}

\newcommand{\WPProxWarmPPool}{= 0.52}

\newcommand{\WPProxAccRDtoR}{0.19}

\newcommand{\WPProxAccPDtoR}{= 0.003}

\newcommand{\WPProxAccRRtoD}{-0.04}

\newcommand{\WPProxAccPRtoD}{= 0.55}

\newcommand{\WPProxIntAccB}{-0.20}

\newcommand{\WPProxIntAccSE}{0.07}

\newcommand{\WPProxIntAccP}{= 0.007}

\newcommand{\WPAccWarmRDtoR}{0.22}

\newcommand{\WPAccWarmPDtoR}{< .001}

\newcommand{\WPAccWarmRRtoD}{-0.01}

\newcommand{\WPAccWarmPRtoD}{= 0.89}

\newcommand{\WPMedA}{0.19}

\newcommand{\WPMedASE}{0.06}

\newcommand{\WPMedAP}{< .001}

\newcommand{\WPMedB}{0.22}

\newcommand{\WPMedBSE}{0.08}

\newcommand{\WPMedBP}{= 0.007}

\newcommand{\WPMedCp}{-0.00}

\newcommand{\WPMedCpP}{= 0.94}

\newcommand{\WPMedInd}{0.041}

\newcommand{\WPMedIndLow}{0.008}

\newcommand{\WPMedIndHigh}{0.086}

\newcommand{\WPMedIndP}{= 0.04}

\newcommand{\MedDoseAccB}{0.08}

\newcommand{\MedDoseAccP}{= .10}

\newcommand{\MedAccWarmthB}{3.46}

\newcommand{\MedAccWarmthP}{< .001}

\newcommand{\MedIndirectB}{0.26}

\newcommand{\MedIndirectLo}{-0.05}

\newcommand{\MedIndirectHi}{0.65}

\newcommand{\MedIndirectP}{= .14}

\newcommand{\MedN}{248}

\newcommand{\DisconfirmPctDtoR}{46}

\newcommand{\MedIndirectBRtoD}{-0.03}

\newcommand{\MedIndirectLoRtoD}{-0.14}

\newcommand{\MedIndirectHiRtoD}{0.08}

\newcommand{\MedNRtoD}{252}

\newcommand{\AbstractTotalN}{%
  \num[group-separator={,},group-minimum-digits=4]{%
    \the\numexpr\WYRN+\WPN+\ExpN+\BehPreN+\LongOneN\relax}}

\begin{document}

\twocolumn[%
  \vspace{4pt}%
  \begin{center}
    {\fontsize{16}{18}\sffamily\bfseries
     Synthetic Contact with AI Reduces Cross-Partisan Animosity\par}
    \vspace{8pt}
    {\sffamily\normalsize
      Benjamin Lira Luttges\textsuperscript{1,\,$\ast$},\enskip
      Noah Castelo\textsuperscript{2},\enskip
      Stefano Puntoni\textsuperscript{1},\enskip
      Olivier Toubia\textsuperscript{3}\par}
    \vspace{4pt}
    {\sffamily\footnotesize
      \textsuperscript{1}The Wharton School, University of Pennsylvania.\enskip
      \textsuperscript{2}Alberta School of Business, University of Alberta.\enskip
      \textsuperscript{3}Columbia Business School, Columbia University.\par
      \vspace{2pt}
      \textsuperscript{$\ast$}Corresponding author: \href{mailto:blira@upenn.edu}{blira@upenn.edu}.\par}
    \ifcameraready
      \vspace{3pt}
      {\footnotesize\sffamily Most recent version: \href{https://lirabenjamin.github.io/synthetic-contact/main-paper.pdf}{lirabenjamin.github.io/synthetic-contact}}%
    \fi
  \end{center}
  \vspace{0.6em}%
  \begingroup
  \leftskip=0.6in \rightskip=0.6in
  \small\sffamily\noindent\textbf{Abstract.\enspace}%
  
Americans' warmth toward members of the opposing political party has fallen sharply over the past three decades---yet meaningful cross-partisan contact remains scarce, in part because people actively avoid it.
Across five preregistered studies (total $N = \AbstractTotalN{}$ U.S.\ partisans), we test whether brief conversations with AI chatbots representing the political outgroup can substitute for the contact people shun.
Synthetic contact first lowers the barrier to entry: partisans would endure almost twice as long contemplating their own mortality to avoid a human outgroup partner as an AI one.
These conversations then correct the misperceptions that fuel division.
At baseline, Democrats placed Republicans more than a standard deviation past their actual position on environmental consumption attitudes---enough to flip the average Republican from supportive to opposed---and a single ten-minute conversation with an outgroup chatbot corrected those beliefs and warmed affect in a within-person study of both parties.
A three-arm experiment ruled out pure engagement and sociality as drivers.
Synthetic contact also moved behavior, in a sample of both parties and on a more affectively charged issue: participants who spoke with an outgroup bot about immigration were six percentage points more likely than controls to choose to have a real conversation with a partisan from the other side.
A final study tested whether these gains last: the warmth effect replicated immediately in a new sample; most of it faded within a week, with a small residual concentrated among the most extreme partisans.
Analyzing conversation content showed that information, more than friendliness, distinguishes outgroup bots from control chatbots.
Together, these findings establish synthetic contact as a scalable, behaviorally consequential, and---unlike face-to-face contact---widely acceptable form of cross-partisan engagement.

  \par
  \endgroup
  \vspace{1em}%
]


\begin{table*}[t]
\centering
\caption{\textbf{Overview of the five studies.} Studies are numbered in the order they appear in the main text. All effect sizes are Cohen's $d$ unless noted; OR denotes an odds ratio. The Pre-registration column lists AsPredicted numbers.}
\label{tab:study_overview}
\small
\begin{tblr}{
  width=\textwidth,
  colspec={Q[l,m]X[1.5,l,m]Q[c,m]X[1.9,l,m]X[2.1,l,m]X[1.7,l,m]X[2.1,l,m]Q[c,m]},
  row{1}={font=\bfseries\sffamily},
  rowsep=3pt,
}
\toprule
No. & Study & $N$ & Design & Key comparison & Primary DV & Key result & Pre-registration \\
\midrule
S1 & Aversion & \WYRN{} & Between-subjects, 2 arms (both parties) & AI bot vs.\ live human outgroup partner & Mortality-reflection minutes accepted at indifference & Bot \WYRMbot{} vs.\ human \WYRMhuman{} min ($d=\WYRd{}$, $p\,\WYRp{}$) & \#286{,}575 \\
S2 & Within-person contact & \WPN{} & Within-person pre--post, 1 arm (both parties) & Pre- vs.\ post-conversation with outgroup bot & Outgroup warmth, 0--100 thermometer & $+\WPWarmthB{}$ pts ($d=\WPWarmthD{}$, $p\,\WPWarmthP{}$) & \#264{,}402 \\
S3 & Three-arm experiment & \ExpN{} & Between-subjects, 3 arms (Democrats) & Outgroup bot vs.\ cats/dogs chat vs.\ Space Invaders & Outgroup warmth, 0--100 thermometer & $d=\ExpCatsDogD{}$ / $\ExpInvadersD{}$ vs.\ the two controls ($p\,\ExpCatsDogP{}$) & \#276{,}530 \\
S4 & Behavioral choice & \BehPreN{} & Between-subjects, 2 arms (both parties) & Outgroup bot vs.\ cats/dogs chat & Costly choice: real outgroup conversation vs.\ mortality reflection & $\mathrm{OR}=\BehPreOR{}$, $p$\,\BehPreP{} & \#287{,}002 \\
S5 & Longitudinal & \LongOneN{} & Between-subjects, 2 arms (Democrats); 1-week follow-up & Outgroup bot vs.\ cats/dogs chat & Outgroup warmth, 0--100 thermometer, at 1 week & Immediate $d=\LongOneD{}$; 1-week $d=\LongTwoD{}$ (n.s., $p\,\LongTwoP{}$); pooled S3+S5 1-week $d=\PoolG{}$ ($p\,\PoolGP{}$) & \#288{,}592 \\
\bottomrule
\end{tblr}
\end{table*}



Americans increasingly view political opponents with suspicion and dislike.
Warmth toward the political outgroup has fallen steadily over the past three decades.\citep{Iyengar2019}
This animosity is only one face of polarization. Beyond cold feelings toward the outgroup, partisans also hold systematic \emph{misperceptions}---inaccurate beliefs about what the other side actually thinks.\citep{Ahler2018, Yudkin2019, MooreBerg2020, Lees2020}
Together these trends erode trust in institutions and make cross-partisan cooperation harder in domains from public health to consumer markets.\citep{Iyengar2019, Hermann2025}
Warming partisans toward their political outgroup remains a central challenge for social scientists and practitioners alike.

Intergroup contact offers one of the most robust solutions to this animosity.
Since Allport,\citep{Allport1954} decades of research have established that positive interactions between members of different groups reduce prejudice and increase mutual understanding,\citep{Pettigrew2006, Hartman2022} though the strength of this evidence has been debated.\citep{Paluck2019}
Researchers still debate why contact works. A classic meta-analysis \citep{Pettigrew2008} of over 500 effects points to three mediators: contact builds knowledge of the outgroup, lowers intergroup anxiety, and increases empathy. Notably, they find that the two affective routes, anxiety and empathy, outweigh the effects of gaining knowledge.

Recent work extends this framework to politics: bringing Democrats and Republicans together for cross-party discussion reduces affective polarization.\citep{Levendusky2021}
Explicitly debating partisan disagreements, however, does not reliably help---Santoro and Broockman \citep{santoro_promise_2022} found that outpartisans who discussed a shared experience (their ``perfect day'') grew less polarized, whereas those who debated their disagreements did not.

Contact between political outgroups faces an important limitation: most partisans actively avoid interacting with the other side.
Using anonymized smartphone location data from millions of Americans, Chen and Rohla \citep{ChenRohla2018} found that Thanksgiving dinners shared across partisan lines ended earlier than same-party dinners, and liberals and conservatives alike forgo money to avoid hearing opposing opinions.\citep{Frimer2017} This avoidance is partly miscalibrated---people overestimate how unpleasant engaging the other side will feel.\citep{Dorison2019}
Face-to-face interventions also demand logistics and professional facilitation that limit their scale.

Digital platforms opened new possibilities, but online political interaction frequently amplifies conflict rather than reducing it.\citep{Bail2018}
Even interventions explicitly designed to expose users to opposing views can fall flat or backfire: a field experiment on Twitter found that replacing participants' feeds with opposing-leaning feeds boosted engagement without improving self-reported understanding of the other side.\citep{Saveski2022}
Another Twitter intervention that recommended opposing-ideology accounts to follow \emph{reduced} users' willingness to converse with an outgroup member.\citep{Gillani2018}

Large language models might offer a way to warm cross-party relationships while avoiding some of the limitations of human contact and interventions on digital platforms.
A chatbot prompted to represent the political outgroup stands in for a real interlocutor, available on demand. We use the term \emph{synthetic contact} for conversations of this kind.
Because no real person sits on the other side, partisans can engage the views they avoid without the threat that makes contact aversive in person and backfire online.
The possibility that AI might improve intergroup relations has been raised conceptually: Hermann et al.\ \citep{Hermann2025} propose that AI agents could reduce prejudice if \emph{engineered} to be counter-stereotypical, deliberately built to contradict the outgroup's negative stereotype.
We test a more minimal version: rather than engineering the bot to defy stereotypes, we prompt it to represent a \emph{typical} outgroup member.
If partisan stereotypes are exaggerated, then even an accurate portrayal should contradict them, and we test whether that alone is enough to reduce prejudice.

Synthetic contact is also newly feasible, and several lines of evidence suggest it could work.
Large language models can generate realistic representations of diverse political viewpoints,\citep{Park2024, Toubia2025} and participants often engage with them as they would a human conversation partner;\citep{NassMoon2000, Klein2025} meta-analytic evidence already shows that digitally mediated intergroup contact reduces prejudice.\citep{PereiraDaCosta2024}
Recent work also shows that people perceive AI sources as less biased, more informative, and less persuasively intended than human sources, which increases receptiveness to opposing views.\citep{lu_how_2025}
People are not only open to LLMs but persuaded by them: brief LLM dialogue can durably reduce conspiracy beliefs,\citep{Costello2024} shift candidate preferences in real elections,\citep{Lin2025} and out-persuade humans in head-to-head debate.\citep{Salvi2024}

But persuasive power cuts both ways.
Large language models trained on internet text exhibit systematic political leanings,\citep{Santurkar2023, Hartmann2023} so a bot prompted to represent the political outgroup may portray its positions in caricatured rather than calibrated form.
If so, its persuasive force would entrench partisan misperceptions rather than correct them---reinforcing the very stereotypes that contact is meant to dissolve.

Here, we test across five preregistered studies whether brief synthetic contact is acceptable, whether it corrects misperceptions and warms cross-partisan affect, and whether it moves a costly behavioral choice (Table~\ref{tab:study_overview} summarizes the design, sample, and primary outcome of each study).
We begin with an incentive-compatible aversion experiment that trades off cross-partisan conversation against an aversive mortality-reflection task, quantifying how much more willing partisans are to engage with an AI outgroup partner than with a human one.
A within-person study with Democrats and Republicans then tests whether a single ten-minute conversation with an outgroup-representing chatbot objectively corrects misperceptions and warms cross-partisan affect.
A three-arm experiment compares synthetic contact against an active chat control (an equally polite AI conversation about an apolitical topic) and a non-social game control, isolating the causal contribution of outgroup-specific content.
A two-arm behavioral experiment asks whether synthetic contact shifts behavior, not just attitudes: whether participants who just spoke with an outgroup bot are more willing to enter a real cross-partisan conversation than control participants.
A longitudinal experiment tests whether brief synthetic contact produces durable attitude change one week later.
Finally, we code the content of every bot conversation to ask which mechanism sets outgroup bots apart from controls---stereotype-disconfirming information (a cognitive route) or warmth and empathy (an affective route).\citep{Pettigrew2008}

Across these studies we deliberately varied the conversation topic: the misperception and warmth studies used environmental consumption attitudes, where a validated scale (the GREEN measure, \citealp{Haws2014}) quantifies misperception item by item, while the two incentive-compatible studies with real behavioral stakes used immigration, a more affectively charged identity issue and a harder test for contact. Convergent effects across both issues suggest that the effect is not specific to any single topic.

\section*{An AI partner halves the aversion to cross-partisan conversation}


Before synthetic contact can reduce polarization, partisans have to agree to it.
In Study 1 (Preregistered, AsPredicted~\#286{,}575, $N = \WYRN$), we measured how much of an aversive experience they would endure to avoid an outgroup conversation, and whether an AI partner lowers that price.
Participants made repeated forced choices between three minutes of conversation about immigration with a member of the political outgroup and an adjustable duration $X$ of reflection on one's own mortality (a deliberately aversive task).
The duration $X$ was raised whenever a participant picked the conversation and lowered whenever they picked mortality reflection, homing in on the duration at which each participant was indifferent between the two.
Only the conversation partner differed between conditions: in the \emph{human} condition, participants weighed mortality reflection against three minutes with a live participant from the opposing party, whereas in the \emph{bot} condition they weighed it against three minutes with an AI trained to represent a typical outgroup member.
The staircase was incentive-compatible: whichever option a participant chose on the final trial was the option they actually had to complete---a real cross-partisan conversation or a real mortality-reflection exercise---so every choice traded real conversation time against real mortality-reflection time.

\begin{figure}[t]
  \centering
  \includegraphics[width=\columnwidth]{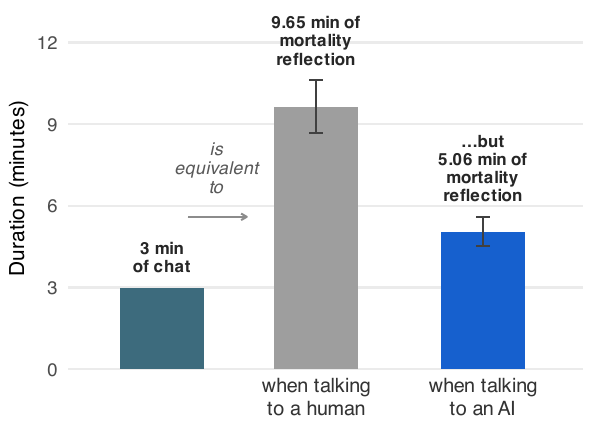}
  \caption{\textbf{Aversion to outgroup conversation is far lower for an AI
           than a human partner.} In an incentive-compatible 1-up-1-down
           staircase, participants traded a fixed three-minute outgroup
           conversation about immigration against an adjustable duration of
           mortality reflection. Bars show the mortality-reflection duration
           that felt equally aversive to the three-minute chat:
           \WYRMhuman{} minutes with a human outgroup partner, but only
           \WYRMbot{} minutes with an AI partner. Error bars are $\pm 1$ SE.}
  \label{fig:wyr}
\end{figure}

Aversion to outgroup conversations was substantially lower when the interlocutor was an AI.
The preregistered primary model regressed the mortality threshold on condition: thresholds were \WYRb{} minutes lower on average in the bot condition than in the human condition ($\beta = \WYRb$, 95\% CI [\WYRciLo, \WYRciHi], $t = \WYRt$, $p \WYRp$, $d = \WYRd$; Figure~\ref{fig:wyr}).
The typical participant matched with a human equated three minutes of outgroup conversation with \WYRMhuman{} minutes of contemplating their own death---roughly double the exchange rate accepted with a bot (\WYRMbot{} minutes).
The advantage held for Democrats and Republicans alike, across the extremity range, and across robustness specifications (Appendix~\ref{app:wyr_mlm}).

Why do partisans refuse a conversation with a member of the other party?
A long literature suggests one likely answer: \emph{they misperceive what the conversation would be like.} Partisans systematically hold exaggerated views of the outgroup's positions, its composition, and its attitudes toward their own side,\citep{Yudkin2019, Ahler2018, Lees2020, MooreBerg2020} and overestimate how aversive exposure to opposing views will be.\citep{Dorison2019}
This suggests that an intervention that corrects such misperceptions may both warm cross-partisan affect and---over time---erode the very aversion that limits face-to-face contact.
We now turn to whether brief AI conversations can correct these misperceptions, durably warm cross-partisan affect, and change behavior to make human intergroup contact more likely.

\section*{Partisans misperceive each other}

\begin{figure*}[t]
  \centering
  \includegraphics[width=\textwidth]{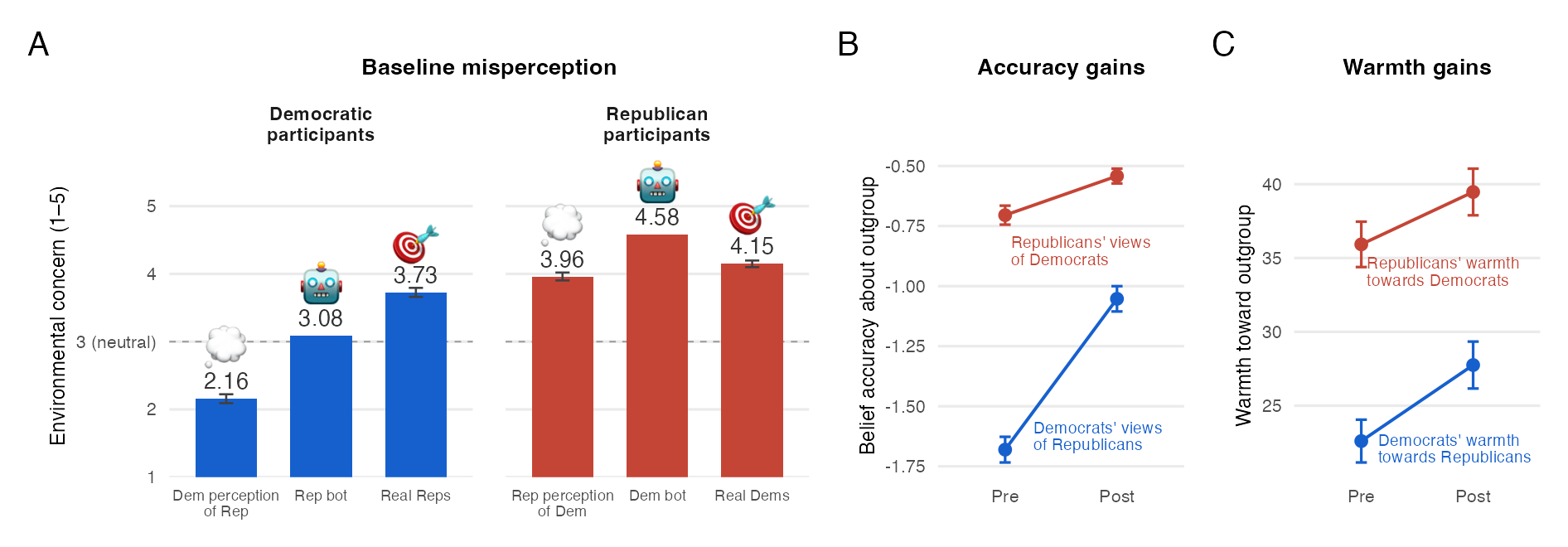}
  \caption{\textbf{A single ten-minute conversation corrects misperceptions of
           the outgroup and warms attitudes toward it.} Color denotes the
           \emph{participant} party (blue = Democrats, red = Republicans).
           \textbf{(A) Baseline misperceptions:} each party's estimate of the
           outgroup, the outgroup's actual attitudes, and the bot's position on
           the six environmental items. Democrats sharply underestimate
           Republican environmental concern, and the bot sits closer to real
           Republicans than Democrats' estimate does.
           \textbf{(B) Belief accuracy} and \textbf{(C) outgroup warmth},
           pre- and post-chat, by party; both improved after the conversation,
           and the size of belief correction predicts the size of warmth gain.
           Error bars are $\pm 1$ SE.}
  \label{fig:wp_main}
\end{figure*}

In Study 2 (Preregistered, AsPredicted~\#264{,}402, $N = \WPN{}$), we asked partisans (\WPNDtoR{} Democrats, \WPNRtoD{} Republicans) to report their own attitudes on six environmental items (the GREEN consumption-values scale, 1--5, \citealp{Haws2014}) and to estimate how a typical outgroup member would respond to the same items.
Partisans held large, asymmetric misperceptions of each other's environmental attitudes (Figure~\ref{fig:wp_main}A).
Democrats sharply underestimated Republican attitudes toward environmental consumption ($d = \WPDemBeliefRealD{}$, SE $= \WPDemBeliefRealSE{}$, $p \WPDemBeliefRealP{}$)---an error large enough to misclassify the average Republican as opposed to green products rather than moderately supportive of them. Republicans were more calibrated about Democrats ($d = \WPRepBeliefRealD{}$, SE $= \WPRepBeliefRealSE{}$, $p \WPRepBeliefRealP{}$). The two misperceptions differ: the belief--reality gap is larger in the Democrat-judging-Republican direction than in the reverse (role $\times$ target interaction, $t(\WPMispercDiffDf{}) = \WPMispercDiffT{}$, $p \WPMispercDiffP{}$).

The bots themselves were imperfect guides: both held more extreme views than the partisans they represented.
To recover the bots' own attitudes, we presented each party-conditioned bot with the same six GREEN items and computed its position on each item as the expected response value, weighting the integers 1--5 by the model's token probabilities at the first response position.\citep{Santurkar2023}
The Republican-representing bot scored \WPRepBot{}, below the Republican mean (\WPRepReal{}) by $d = \WPRepBotRealD{}$; the Democrat-representing bot scored \WPDemBot{}, above the Democratic mean (\WPDemReal{}) by $d = \WPDemBotRealD{}$.
Which direction the bot erred in mattered more than that it erred at all, because what helps a learner is a guide closer to the truth than their own starting beliefs.
For Democrats, who began badly miscalibrated about Republicans, the Republican bot was a far more accurate guide than their own beliefs, so leaning toward it moved them toward the truth.
For Republicans, who began less misaligned about the outgroup than Democrats, the bot offered less to correct---though they still became more accurate (Fig.~\ref{fig:wp_main}B).
This asymmetry anticipates the accuracy gains reported next: the bot had much more to teach Democrats than Republicans.

\section*{Brief synthetic contact corrects misperceptions and warms cross-partisan affect}


The same \WPN{} participants then conversed for ten minutes with a chatbot prompted to represent their political outgroup, with each participant instructed to learn how the outgroup thinks about environmental policy. The chatbot (GPT-4o) received a brief system prompt instructing it to answer as a typical member of the participant's outgroup, specifying the outgroup's party identity but no scripted positions (full prompts in Appendix~\ref{app:system_prompts}). After the conversation, they re-estimated outgroup attitudes on the same six items and re-rated their warmth toward the outgroup on a 0--100 thermometer.

A single ten-minute conversation corrected baseline misperceptions and warmed cross-partisan affect (Figure~\ref{fig:wp_main}B--C).
Belief accuracy improved by \WPAccB{} points on the 5-point scale ($p \WPAccP{}$, $d = \WPAccD{}$).
Outgroup warmth rose by \WPWarmthB{} thermometer points ($p \WPWarmthP{}$, $d = \WPWarmthD{}$)---equivalent to reversing \WPAncovaYearsLong{} years of the rising partisan animosity documented in national surveys since the 1970s.\citep{Iyengar2019}
Both parties gained, but the gains were graded by baseline misperception.
Democrats, who started with larger misperceptions, gained more in accuracy ($d_{\text{D}\to\text{R}} = \WPAccDDtoR{}$) than Republicans ($d_{\text{R}\to\text{D}} = \WPAccDRtoD{}$; interaction $b = \WPAccIntB{}$, $p \WPAccIntP{}$).
The same gradient held for warmth (Democrats $d = \WPWarmthDDtoR{}$; Republicans $d = \WPWarmthDRtoD{}$).

The size of the accuracy gain tracked the size of the warmth gain.
Controlling for both baseline accuracy and baseline warmth, post-chat belief accuracy strongly predicted post-chat outgroup warmth ($b = \WPAccWarmthB{}$, 95\% CI [\WPAccWarmthCiLo{}, \WPAccWarmthCiHi{}], $p \WPAccWarmthP{}$): participants whose beliefs about the outgroup moved more during the conversation also warmed up more, the pattern a cognitive account predicts.

Participants' own perceptions pointed the same way. With both ratings entered together, the bot's perceived informativeness was the only significant predictor of who warmed most ($b = \ExpInfoWarmthB{}$, $p \ExpInfoWarmthP{}$); perceived empathy was not ($b = \ExpEmpWarmthB{}$, $p \ExpEmpWarmthP{}$), though the two coefficients did not differ reliably from each other ($\Delta b = \ExpInfoEmpDiffB{}$, $p \ExpInfoEmpDiffP{}$). The conversations themselves carried the same signal: those delivering more stereotype-disconfirming substance produced the largest belief corrections (Appendix~\ref{app:content_dose}).


\section*{Synthetic contact outperforms two active controls}

\begin{figure}[t]
  \centering
  \includegraphics[width=\linewidth]{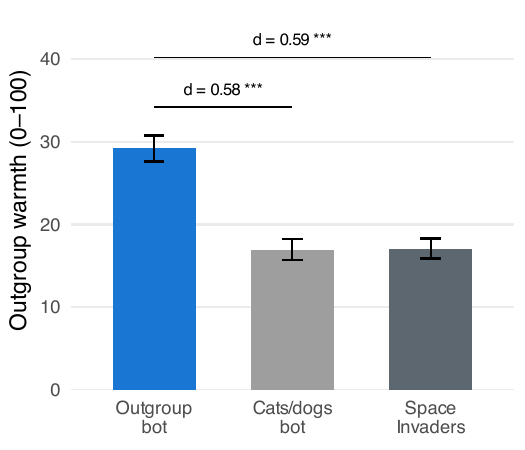}
  \caption{\textbf{Synthetic contact raised outgroup warmth above both
           controls.} Mean outgroup warmth (0--100 thermometer) by condition
           in the three-arm experiment (Study 3). Error bars are $\pm 1$ SE.}
  \label{fig:exp_warmth}
\end{figure}


In Study 3 (Preregistered, AsPredicted~\#276{,}530, $N = \ExpN{}$), we randomly assigned Democrats to one of three conditions: synthetic contact (a ten-minute conversation with a chatbot prompted to represent a Republican, in which participants posed as policy researchers gauging how Republicans think about environmental policy), a chat control (the same interface, but a bot given no political identity and prompted to debate cats vs.\ dogs), or a game control (Space Invaders).
The two controls isolate different confounds: the chat control holds the social, conversational experience constant while stripping outgroup content, reducing the plausibility of a pure-sociality account, whereas the game control removes conversation entirely, ruling out generic engagement or arousal.

Synthetic contact increased outgroup warmth relative to both control conditions (Fig.~\ref{fig:exp_warmth}).
Participants who conversed with the AI outgroup representative rated Republicans more warmly ($M = \ExpSCM{}$, $SD = \ExpSCSD{}$) than those in the cats-and-dogs chat condition ($M = \ExpCatsDogM{}$, $SD = \ExpCatsDogSD{}$; $d = \ExpCatsDogD{}$, $p \ExpCatsDogP{}$) and those in the Space Invaders condition ($M = \ExpInvadersM{}$, $SD = \ExpInvadersSD{}$; $d = \ExpInvadersD{}$, $p \ExpInvadersP{}$).
The two control conditions did not differ from each other ($d = \ExpCatsDogInvadersD{}$, $p \ExpCatsDogInvadersP{}$), confirming that the effect is specific to outgroup-relevant conversation rather than to the experience of chatting with an AI or engaging in an unrelated task.

\section*{Synthetic contact moves a costly behavioral choice}

\begin{figure*}[t]
  \centering
  \includegraphics[width=0.92\textwidth]{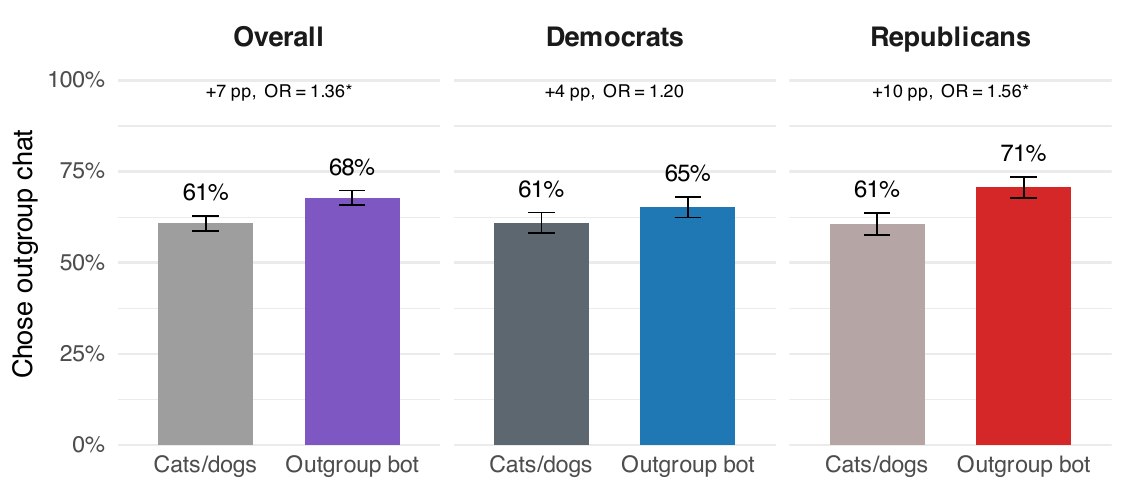}
  \caption{\textbf{After synthetic contact, more partisans choose a real
           cross-partisan conversation} ($N = \BehPreN{}$).
           Pooled choice share (left) and by party
           (Democrats, center; Republicans, right).
           The outcome is the share of participants choosing a three-minute
           conversation with a real outgroup member over a three-minute
           mortality reflection. Error bars are $\pm 1$ SE.}
  \label{fig:behavioral_mt3}
\end{figure*}


In Study 4 (Preregistered, AsPredicted~\#287{,}002, $N = \BehPreN{}$), we randomly assigned Democrats and Republicans to a five-minute chat with an outgroup bot (discussing how the outgroup thinks about immigration) or the cats-and-dogs control, then offered an incentive-compatible binary choice: a three-minute conversation with a real member of their political outgroup, or three minutes on an aversive mortality-reflection task. Whichever option a participant chose, they actually completed.

Synthetic contact shifted a costly behavioral choice toward cross-partisan contact. \BehPrePctCats{} of participants in the cats-and-dogs control chose the outgroup conversation, compared with \BehPrePctBot{} in the outgroup-bot condition.
A preregistered logistic regression with party and mean-centered political extremity as covariates yielded an odds ratio of \BehPreOR{} (95\% CI [\BehPreORlo{}, \BehPreORhi{}], $p$ \BehPreP{}); a Wilcoxon rank-sum robustness test reached the same conclusion ($W = \BehPreW{}$, $p$ \BehPreWP{}).
Republicans showed a significant shift ($\text{OR} = \BehPreRepOR{}$, $p$ \BehPreRepP{}), and Democrats a directional but non-significant one ($\text{OR} = \BehPreDemOR{}$, $p$ \BehPreDemP{}); the two parties did not significantly differ (treatment $\times$ party interaction $\text{OR} = \BehPrePartyInterOR{}$, 95\% CI [\BehPrePartyInterCiLo{}, \BehPrePartyInterCiHi{}], $p$ \BehPrePartyInterP{}).
We found no statistically significant evidence of moderation by partisan strength (treatment $\times$ mean-centered extremity interaction $\text{OR} = \BehPreInterOR{}$, $p$ \BehPreInterP{}; median-split $\text{OR}_{\text{moderate}} = \BehPreModOR{}$ vs.\ $\text{OR}_{\text{strong}} = \BehPreStrongOR{}$).

Synthetic contact does not just shift self-reported warmth: it moves a subsequent costly choice that leads to an interaction with a real outgroup member. The effect also replicates at half the dosage of the earlier warmth experiments---participants here chatted for five minutes rather than the ten used in the three-arm and within-person studies---suggesting even five minutes suffice to shift behavior.

\section*{Most of the warmth effect fades within a week; a small residual concentrates among extreme partisans}


\begin{figure*}[!t]
  \centering
  \includegraphics[width=0.8\textwidth]{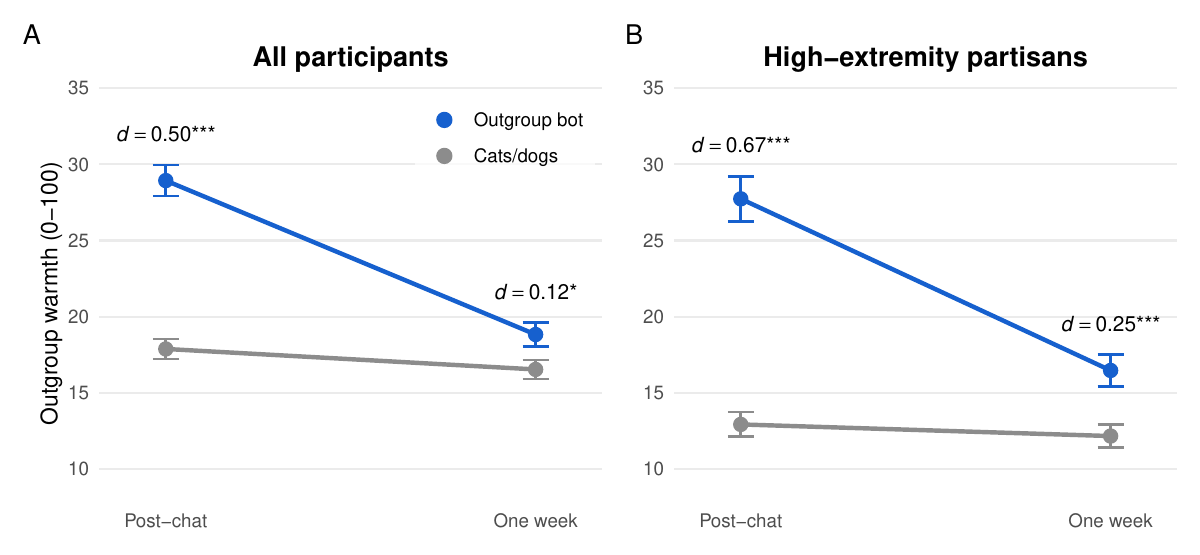}
  \caption{\textbf{Synthetic contact's effect on outgroup warmth is large
           immediately, mostly gone within a week, and larger among extreme
           partisans.} Outgroup warmth immediately after the chat and one week
           later, by condition, pooled across the three-arm follow-up and the
           longitudinal study ($N = \PoolN{}$), for \textbf{(A)} all
           participants and \textbf{(B)} the more extreme half. Points are
           study-adjusted means $\pm 1$~SE; annotations give the pooled effect
           (Cohen's $d$) of the outgroup bot versus control at each timepoint,
           so the plotted gap equals the annotated effect (stars:
           $^{*}p<.05$, $^{***}p<.001$; Appendix~\ref{app:pooled_persistence}).}
  \label{fig:longitudinal_trajectory_main}
\end{figure*}

In Study 5 (Preregistered, AsPredicted~\#288{,}592, $N = \LongOneN{}$), we randomly assigned self-identified Democrats to a five-minute chat with either the outgroup bot (discussing how Republicans think about environmental policy) or the cats-and-dogs control bot, then re-contacted them one week later for a follow-up measure of outgroup warmth.
Relative to attitudes built over years and presumably reinforced daily by media exposure, a five-minute conversation is a light intervention, so we expect substantial decay, consistent with other brief depolarization treatments.\citep{santoro_promise_2022, Voelkel2024}
The goal of this follow-up is therefore to replicate the immediate effect at larger scale and bound how much of it survives a week.
Immediately after the chat, the outgroup bot raised post-chat warmth above the cats-and-dogs control by $b = \LongOneB{}$ points ($t(\LongOneDF{}) = \LongOneT{}$, $p \LongOneP{}$; $d = \LongOneD{}$), replicating the immediate effect in a sample 1.6 times the size of the three-arm experiment.

One week later, $\LongTwoN{}$ of $\LongOneN{}$ participants returned ($\LongAttrAllRate\%$), with no differential attrition by condition (Appendix~\ref{app:longitudinal_ipaw}).
Most of the immediate effect had faded.
The one-week warmth distribution is strongly floor-bunched (Appendix~\ref{app:long_rank_robust}), which violates the assumptions of the preregistered mean-difference ANCOVA; that model returns a small, non-significant residual ($b = \LongTwoB{}$, 95\% CI $[\LongTwoCiLo{}, \LongTwoCiHi{}]$, $d = \LongTwoD{}$, $p \LongTwoP{}$).
A rank-based ANCOVA---a small deviation from the preregistered Wilcoxon rank-sum robustness test that simply adds baseline control---shows the effect declining from $d_{\text{rank}} = \LongOneRankD{}$ immediately after the chat to $d_{\text{rank}} = \LongTwoRankD{}$ one week later, about $\LongRankSurvivePct{}\%$ surviving (a $\LongTwoRankPct{}$-percentile-point shift, 95\% CI $[\LongTwoRankPctCiLo{}, \LongTwoRankPctCiHi{}]$, $p \LongTwoRankP{}$).
Re-fitting the paper's other warmth contrasts with the same rank estimator leaves their conclusions unchanged (Appendix~\ref{app:long_rank_robust}, Table~\ref{tab:rank_robustness}).

The surviving residual warmth is concentrated among the most extreme partisans (high-extremity one-week effect: $\LongTwoRankUpperPct{}$ percentile points, $p \LongTwoRankUpperP{}$); however, the preregistered extremity interaction is not significant ($b = \LongTwoExtIntB{}$, $p \LongTwoExtIntP{}$).

We also recontacted participants from the three-arm experiment (Study 3) one week later ($N = \PoolNExp{}$), and its residual effect matches the longitudinal study's almost exactly (Cohen's $d = \PoolExpG{}$ vs.\ $d = \PoolLongG{}$; the effect does not differ between the two studies, condition~$\times$~study $p \PoolHetP{}$). Pooling both follow-ups in an exploratory individual-participant analysis ($N = \PoolN{}$), the one-week residual is significantly positive ($d = \PoolG{}$, 95\% CI $[\PoolGciLo{}, \PoolGciHi{}]$, $p \PoolGP{}$) and concentrated among the more extreme half of partisans ($d = \PoolHiG{}$ $[\PoolHiCiLo{}, \PoolHiCiHi{}]$, $p \PoolHiP{}$; Appendix~\ref{app:pooled_persistence}). A complementary Bayesian analysis with weakly informative priors places a $\PoolBrmsProbSc{}$ posterior probability on a positive one-week effect (Appendix~\ref{app:pooled_persistence}). A single brief conversation thus leaves a faint but consistent week-later trace, strongest among the most extreme partisans.

This decay parallels other brief depolarization interventions: the effects of single cross-partisan conversations had faded by a three-month follow-up,\citep{santoro_promise_2022} and brief online treatments decayed substantially within two weeks.\citep{Voelkel2024}

\section*{Outgroup bots differ from control bots most in what they say, not how warmly they say it}

    
\begin{figure*}[!t]
  \centering
  \includegraphics[width=0.95\textwidth]{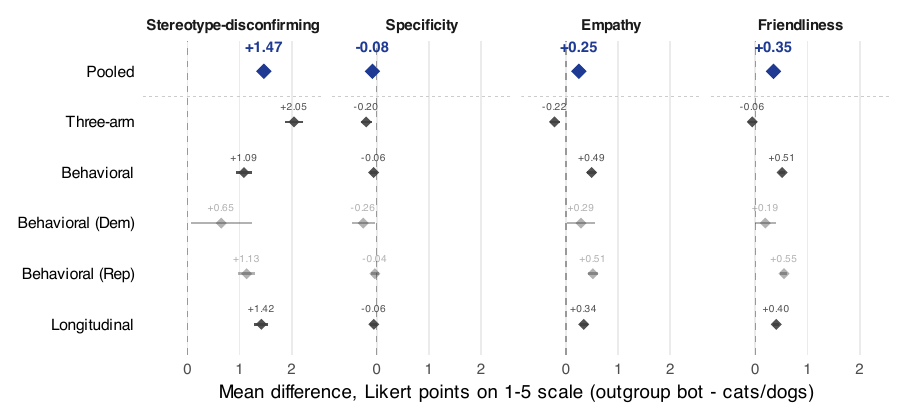}
  \caption{\textbf{Outgroup bots differ from control bots more in information
           than in friendliness.}
           Per-study and pooled mean differences (outgroup bot $-$
           cats-and-dogs control) in GPT-5.4-mini's ratings of each
           conversation on four dimensions (1--5 Likert), across the
           three-arm, behavioral, and longitudinal studies; the behavioral
           contrast is party-adjusted and subgroup estimates are faded.
           Error bars are 95\% CIs; the pooled diamond is the
           inverse-variance-weighted average. The cognitive-route difference
           (stereotype-disconfirming substance) is the largest and most
           consistent contrast, present in every study and reliably larger
           than the affective-route difference
           (Appendix~\ref{app:route_interaction}).}
  \label{fig:cross_study_process}
\end{figure*}

To characterize what the outgroup bots said to participants, and how they said it, we audited every conversation across studies in an exploratory, non-preregistered analysis.
We used GPT-5.4-mini to score all $\ConvAllN{}$ conversations on four dimensions on a $1$--$5$ scale: how much the bot delivered \emph{stereotype-disconfirming substance} and \emph{informational specificity} (the cognitive route to prejudice reduction), and how much \emph{empathy} and \emph{friendliness} it conveyed (the affective route).\citep{Pettigrew2008, Hermann2025}
Figure~\ref{fig:cross_study_process} shows the resulting condition contrasts in the three studies with a control arm; coding and analysis details are in Methods.

More than anything else, the outgroup bots offered stereotype-disconfirming substance: positions that cut against what their partner expected from the other side.
Pooled across the three studies with a control arm, outgroup bots scored $M = \ConvPooledDiscOutM{}$ on stereotype-disconfirming substance versus $M = \ConvPooledDiscCatsM{}$ for cats-and-dogs controls (1--5 scale), and the per-study gaps were consistently large: $\Delta M = +\ConvExpDiscDiff$ points in the three-arm experiment ($p$ \ConvExpDiscP{}), $\Delta M = +\ConvBehDiscDiff$ points in the behavioral study ($p$ \ConvBehDiscP{}), and $\Delta M = +\ConvLongDiscDiff$ points in the longitudinal study ($p$ \ConvLongDiscP{}).
This was not because the outgroup bots simply conveyed more detailed information: on informational specificity the control bots were, if anything, marginally higher within their own topics ($\Delta M = \ConvExpSpecDiff$ points in the experiment, $p$ \ConvExpSpecP{}; $\ConvBehSpecDiff$ behavioral; $\ConvLongSpecDiff$ longitudinal), so the cognitive contrast came from \emph{what} the outgroup bot talked about, not how granular it was.

The affective dimensions tell a less consistent story.
In the three-arm experiment, GPT rated the cats-and-dogs bot as \emph{more} empathic than the outgroup bot ($\Delta M = \ConvExpEmpDiff$ points, $p$ \ConvExpEmpP{}) and roughly as friendly ($\Delta M = \ConvExpWarmDiff$ points, $p$ \ConvExpWarmP{}): the bot that produced the warmth gain was, if anything, the less pleasant of the two.
In the behavioral and longitudinal studies, however, GPT rated the outgroup bot as both more empathic ($\Delta M = +\ConvBehEmpDiff$, $p$ \ConvBehEmpP{}; $\Delta M = +\ConvLongEmpDiff$, $p$ \ConvLongEmpP{}) and friendlier ($\Delta M = +\ConvBehWarmDiff$, $p$ \ConvBehWarmP{}; $\Delta M = +\ConvLongWarmDiff$, $p$ \ConvLongWarmP{}) than the chat control.

Outgroup and control bots differed more in stereotype-disconfirming content than in empathy or friendliness.
Pooled across studies, the outgroup-vs-control gap was $\RouteIntB$ Likert points wider on the cognitive dimensions than on the affective ones (95\% CI $[\RouteIntCiLo, \RouteIntCiHi]$, $p$ \RouteIntP{}), and the same ordering held in every individual study (Methods; Appendix~\ref{app:route_interaction}).

Representative verbatim excerpts of stereotype-disconfirming bot messages appear in Appendix~\ref{app:disconfirm_excerpts}.

\section*{Discussion}


Brief AI synthetic contact is perceived as more acceptable than face-to-face contact, corrects misperceptions, warms cross-partisan affect, moves a costly behavioral choice, and---like other brief contact interventions---attenuates within a week.

Human intergroup contact works mainly through an affective route; synthetic contact appears to work mainly through a cognitive one---the affective channel was present in our conversations too, but consistently smaller (Fig.~\ref{fig:cross_study_process}). The Pettigrew \& Tropp \citep{Pettigrew2006, Pettigrew2008} meta-analysis of $> 500$ contact studies finds that the dominant mediators of human contact effects are affective---reduced intergroup anxiety and increased empathy---rather than purely cognitive. Consistent with that affective-mechanism account, Santoro and Broockman \citep{santoro_promise_2022} found that randomly assigning outpartisan strangers to discuss a shared experience reduced affective polarization, but assigning them to discuss \emph{partisan disagreement} did not: explicit political topics appear to introduce anxiety and threat that erode the affective gains human conversation otherwise produces. 

We find the opposite pattern: explicit political disagreement, which fails to warm relations between human partisans, warmed them when the interlocutor was an AI. The most plausible reason is that an AI partner removes the interpersonal stakes that make political talk aversive---there is no risk of being judged and no face to manage---so participants engage with the disagreement on its merits rather than bracing against it. Freed of that threat, participants could take in the bot's stereotype-disconfirming points---the process our content analyses point to, though we did not manipulate it directly. This evidence is exploratory, but it raises a possibility: the advantage of synthetic contact over face-to-face contact may not be only that it scales. By replacing the person on the other side with a bot, it may make the disagreements that matter most easier to discuss.

The introduction raised the risk that politically slanted language models would caricature the outgroup and entrench misperceptions. That risk partly materialized---both bots held more extreme positions than the partisans they represented---yet beliefs became more accurate, not less, because a guide need not be perfect to help, only less wrong than the learner. Bots calibrated against real survey data should do better still.

As expected for so brief an intervention, most of a single five-minute conversation's effect faded within a week, though a small residual persisted among the most extreme partisans---the group these interventions most need to reach. The strength of synthetic contact is that it can be repeated: unlike a face-to-face meeting, a bot conversation is available again whenever and wherever partisans are already online. Even the immediate effect carries weight---right after the conversation, partisans were more likely than controls to choose a real outgroup exchange. Whether repeated synthetic contact can build lasting warmth is a question for future work.

Two limitations qualify these findings. First, we indexed affective polarization mainly as outgroup warmth on a feeling thermometer---the standard measure, but one facet of a multidimensional construct \citep{Druckman2019}---and did not test social distance, trait attributions, or support for anti-democratic action. Second, the evidence comes from online U.S.\ partisans and two issues, environment and immigration, so generalization across populations and topics remains open. The belief-accuracy gains, observed in a within-person design, also warrant more caution than the experimentally isolated warmth effects.

Traditional contact interventions work \citep{Pettigrew2006} but do not scale, largely because people will not accept them.\citep{Frimer2017}
Synthetic contact reverses that constraint: it is inexpensive, available on demand, and far more acceptable than the face-to-face contact partisans avoid.
Talking to an outgroup-representing bot corrects misperceptions and warms cross-partisan affect.
Whether the effects shown here are sufficient to matter at the societal level remains an open question, but the cost-effectiveness of synthetic contact warrants optimism.
A brief chatbot interaction embedded in a news app, social media platform, or civic engagement tool could reach millions of partisans where they already are: online, often alone, and unwilling to enter a room with the other side.

\ifcameraready
  \balance 
\fi        
\section*{Methods}

\subsection*{Aversion study}


\paragraph*{Participants.}
We recruited partisans via CloudResearch Connect with a preregistered target of 500 completions; owing to a recruitment-setting error, the study was fielded for 600, yielding \WYRN{} eligible participants (\WYRNdem{} Democrats, \WYRNrep{} Republicans); results are unchanged if the sample is restricted to the first 500 respondents (Section~\ref{app:wyr_mlm}). Random assignment yielded \WYRNbot{} participants in the bot condition and \WYRNhuman{} in the human condition. The study was preregistered at AsPredicted~\#286{,}575.

\paragraph*{Design.}
Two between-subjects conditions differed only in the identity of the conversation partner. In the \emph{bot} condition, participants conversed for three minutes with an AI trained to represent a typical member of the opposing party. In the \emph{human} condition, participants were paired in real time with another live participant from the opposing party for a three-minute text conversation. Topic (immigration), conversation length, staircase mechanics, and the alternative task were identical across conditions.

\paragraph*{Procedure.}
An adaptive 1-up-1-down staircase estimated each participant's indifference point: the duration of mortality reflection equated to a three-minute outgroup conversation. On each trial, participants chose between ``three minutes of conversation with an outgroup partisan / AI about immigration'' and ``$X$ minutes of mortality reflection.'' $X$ started at 10 minutes with a halving step size (5 $\to$ 2.5 $\to$ 1.25 $\to$ 0.625 $\to$ 0.25 minutes) and a floor of 0.25 minutes. The staircase terminated after 12 trials, after three small reversals (step $\leq 0.5$), or when the same duration was shown in two consecutive trials with the same choice. The threshold is the mean of the last two reversals, or the last-shown duration if fewer than two reversals occurred. The design was incentive-compatible: whichever option participants picked on the final trial, they actually had to complete.

\paragraph*{Exclusions.}
Per preregistration, we excluded participants who did not identify as either a Democrat or a Republican.

\paragraph*{Analytic strategy.}
The primary preregistered analysis regresses the mortality threshold on condition: \texttt{threshold $\sim$ condition}. Preregistered secondary analyses add a condition$\times$extremity interaction and a party control. Robustness checks include a Mann--Whitney test on the raw thresholds and a mixed-effects logistic regression on trial-level choices (\texttt{chose\_chat $\sim$ duration * condition + (1 + duration | participant)}); see Appendix~\ref{app:wyr_mlm}.

\subsection*{Within-person study}


\paragraph*{Participants.}
We recruited \WPN{} participants via Prolific (\WPNDtoR{} Democrats learning about Republicans; \WPNRtoD{} Republicans learning about Democrats). The study was preregistered at AsPredicted~\#264{,}402.

\paragraph*{Design.}
The study used a within-person pre-post design with a single condition: every participant conversed with a chatbot prompted to represent their political outgroup (Republicans for Democratic learners; Democrats for Republican learners).

\paragraph*{Procedure.}
Participants first rated their warmth toward the political outgroup on a $0$--$100$ feeling thermometer (pre-interaction). They then conversed for ten minutes with the outgroup-representing chatbot, tasked with learning how the outgroup thinks about environmental policy. Immediately afterward, they re-rated outgroup warmth on the same thermometer (post-interaction) and rated the chatbot's informativeness and empathy on $5$-point Likert scales.

\paragraph*{Exclusions.}
The preregistration specified no exclusions from the primary analyses, so all \WPN{} participants are retained.

\paragraph*{Analytic strategy.}
The primary preregistered analysis is a mixed-effects model on the pre-post warmth ratings: \texttt{warmth $\sim$ time * learner\_party + (1 | participant)}, where \texttt{time} contrasts pre and post and \texttt{learner\_party} contrasts D$\to$R and R$\to$D. The coefficient on \texttt{time} captures the average pre-post change; the interaction tests whether the change differs between learner groups. Models were estimated via maximum likelihood with \texttt{lme4} \citep{Bates2015} and Satterthwaite degrees of freedom.\citep{Luke2017}

\subsection*{Three-arm experiment}


\paragraph*{Participants.}
We recruited \ExpNRecruited{} self-identified Democrats via Prolific. The study was preregistered on AsPredicted~\#276{,}530.

\paragraph*{Design.}
Participants were randomly assigned to one of three conditions: (1) \emph{synthetic contact}---a ten-minute conversation with a chatbot prompted to represent a Republican, in which participants role-played a nonpartisan policy researcher tasked with learning how Republicans think about environmental policy; (2) \emph{chat control}---a ten-minute conversation with the same interface, but with a bot instructed to debate whether cats or dogs are better and to avoid political topics, not to represent an outgroup member; or (3) \emph{game control}---a ten-minute session playing Space Invaders, a browser-based arcade game with no conversational component.

\paragraph*{Procedure.}
All participants completed their assigned ten-minute intervention. Immediately afterward, they rated outgroup warmth on a $0$--$100$ feeling thermometer.

\paragraph*{Exclusions.}
Per preregistration, we excluded participants who met any of the following criteria: reCAPTCHA score below $0.5$, more than two tab switches during the intervention, or a total completion duration more than three standard deviations below the sample mean. After exclusions, the analytic sample comprised $N = \ExpN{}$ participants (\ExpNSC{} synthetic contact, \ExpNCatsDogs{} cats-and-dogs chat, \ExpNInvaders{} Space Invaders).

\paragraph*{Analytic strategy.}
The primary preregistered analysis is an OLS regression of outgroup warmth on condition: \texttt{warmth $\sim$ condition}, with outgroup bot as the reference level. The two planned contrasts test our preregistered hypotheses (H1: outgroup bot $>$ cats-and-dogs; H2: outgroup bot $>$ Space Invaders). Wilcoxon rank-sum tests on each contrast provide non-parametric robustness given the zero-inflated distribution of warmth ratings.

\subsection*{Behavioral study}


\paragraph*{Participants.}
We recruited self-identified Democrats and Republicans via Prolific, screened for above-threshold political extremity (excluding participants with extremity $< 10$ on the $0$--$100$ scale). The study was preregistered at AsPredicted~\#287{,}002, which specified a three-minute mortality-reflection comparison task and a sampling frame including Republicans recruited $50/50$ with Democrats. Analyses are restricted to the preregistered sample---participants randomly assigned to the two chat conditions while the preregistration was in force ($N = \BehPreN{}$; \BehPreDemN{} Democrats, \BehPreRepN{} Republicans).

\paragraph*{Design.}
Participants were randomly assigned between subjects to one of two conditions: (1) \emph{outgroup bot}---a five-minute conversation with a chatbot prompted to represent a typical member of their political outgroup, with participants instructed to learn how the outgroup thinks about immigration, or (2) \emph{chat control}---a five-minute conversation with the same interface, but with a bot instructed to debate whether cats or dogs are better and to avoid political topics, not to represent an outgroup member. Both chatbots used the same conversational interface, the same underlying model (GPT-5.2-mini), and the same politeness norms; they differed only in assigned topic.

\paragraph*{Procedure.}
Participants completed their assigned five-minute conversation, then faced an incentive-compatible binary choice between (a) ``Have a three-minute conversation with a real \{Republican/Democrat\}'' (outgroup label matched the participant's party) or (b) ``Complete a three-minute mortality reflection exercise.'' Whichever option they chose, they actually completed it: real cross-partisan conversation vs.\ real mortality reflection. The primary outcome was whether the participant chose the outgroup conversation ($1$) or the mortality reflection ($0$).

\paragraph*{Exclusions.}
Per preregistration, we excluded participants who met any of the following criteria: a Qualtrics reCAPTCHA score below $0.5$, more than two tab switches during the chatbot page, a total completion duration more than three standard deviations below the sample mean, Prolific's bot/LLM detection flagging the submission, or sending no messages to the chatbot.

\paragraph*{Analytic strategy.}
The primary preregistered analysis is a logistic regression: \texttt{task\_choice $\sim$ condition + party + extremity\_c}. We report the condition odds ratio, $95\%$ confidence interval, and two-sided $p$ value. As a non-parametric robustness check we report a Wilcoxon rank-sum test. Party and political extremity are examined as preregistered exploratory moderators.

\subsection*{Longitudinal study}


\paragraph*{Participants.}
We recruited self-identified Democrats (target $N = 1{,}200$ before exclusions) via CloudResearch Connect in three batches launched 2026-05-01, 2026-05-04, and 2026-05-06. Participants were screened on partisan identification and on a $0$--$100$ political-extremity slider; those scoring below $30$ on extremity were screened out before random assignment. The study was preregistered at AsPredicted~\#288{,}592.

\paragraph*{Design.}
Participants were randomly assigned between subjects to one of two conditions: (1) \emph{outgroup bot}---a five-minute conversation with a chatbot prompted to represent a typical Republican, with participants instructed to learn how Republicans think about environmental policy, or (2) \emph{chat control}---a five-minute conversation with the same interface, but with a bot instructed to debate whether cats or dogs are better and to avoid political topics, not to represent an outgroup member. The two conditions are matched on interface, duration, model, and politeness norms, differing only in conversation topic.

\paragraph*{Procedure.}
At Time~1, participants rated outgroup warmth on a $0$--$100$ feeling thermometer (\texttt{warmth\_T1}), completed their assigned five-minute conversation, and rated outgroup warmth again immediately post-chat. One week later, all Time~1 completers were invited back and re-rated outgroup warmth on the same thermometer (\texttt{warmth\_T2}, the primary dependent variable).

\paragraph*{Exclusions.}
Per preregistration, we excluded participants who met any of the following criteria: failure of any of three embedded attention/bot checks (line-instruction, math, animal-recognition), a Qualtrics reCAPTCHA score below $0.5$, more than two tab switches during the chatbot page, a total completion time more than three standard deviations below the sample mean, or sending no messages to the chatbot. After exclusions, $N = \LongOneN{}$ participants were retained at Time~1 ($\LongOneNbot{}$ outgroup bot, $\LongOneNctrl{}$ cats-and-dogs control).

\paragraph*{Analytic strategy.}
The primary preregistered analysis is an ANCOVA: \texttt{warmth\_T2 $\sim$ condition + warmth\_T1}. Preregistered secondary analyses include a Wilcoxon rank-sum test of \texttt{warmth\_T2} by condition, a \texttt{condition * extremity\_c} interaction added to the primary ANCOVA, and the simple effect of condition within the upper half of the extremity distribution (median split). Because the Wave-2 warmth distribution showed strong floor bunching, we additionally report a rank ANCOVA on percentile-transformed outcomes; the attrition-weighting robustness check is in Appendix~\ref{app:longitudinal_ipaw}.

\subsection*{Cross-study content audit}


\paragraph*{Conversations coded.}
We coded all $\ConvAllN{}$ bot conversations from the within-person, three-arm, behavioral, and longitudinal studies. The within-person study has no chat-control arm, so it does not contribute an outgroup-vs-control contrast and is omitted from the forest plot; the three studies with a control arm contribute the contrasts shown in Figure~\ref{fig:cross_study_process}.

\paragraph*{Coding scheme.}
Each conversation was scored on four theoretically motivated dimensions on a $1$--$5$ scale: two indexing the cognitive route to prejudice reduction---\emph{stereotype-disconfirming substance} and \emph{informational specificity}---and two indexing the affective route---\emph{empathy} and \emph{friendliness}.\citep{Pettigrew2008, Hermann2025} To block cross-dimension halo bias, each dimension was coded in a separate API call (GPT-5.4-mini) with no information about the other dimensions; the model wrote one-sentence reasoning before scoring.

\paragraph*{Per-dimension contrasts.}
For each dimension we computed the outgroup-vs-control mean difference, $95\%$ CI, and Cohen's $d$ within each study. The behavioral contrast is adjusted for participant party because random assignment of bot prompt is crossed with party in that study; the other contrasts are unadjusted.

\paragraph*{Cognitive-vs-affective comparison.}
To test whether the condition contrast is larger on the cognitive dimensions than on the affective ones, we pivoted the per-conversation scores to long format (one row per dimension) and fit a linear mixed-effects model: \texttt{score $\sim$ condition * route + (1 | participant)}, where \texttt{route} groups the two cognitive dimensions versus the two affective ones. A positive condition$\times$route interaction means the outgroup-vs-control gap is wider on the cognitive route than on the affective route. We fit this model separately within each of the three studies with a control arm and pooled across all three (Appendix~\ref{app:route_interaction}, Table~\ref{tab:route_interaction}).

\paragraph*{Preregistration.}
This content audit was not preregistered. It is an exploratory characterization of what the bots said and how they said it, not a confirmatory test.

\subsection*{Ethics}

All studies were approved by the University of Pennsylvania Institutional Review Board (protocol~\#860019). All participants provided informed consent before participation.

\subsection*{Data and code availability}


\noindent All de-identified data and analysis code are available at Zenodo (\url{https://doi.org/10.5281/zenodo.20971465}).

\subsection*{Author contributions, competing interests, and funding}

\noindent\textbf{Author contributions.} B.L.L.: conceptualization, data curation, formal analysis, funding acquisition, investigation, methodology, resources, software, validation, visualization, writing (original draft), and writing (review and editing). N.C.: conceptualization, software, visualization, and writing (review and editing). S.P.: conceptualization, funding acquisition, supervision, visualization, and writing (review and editing). O.T.: conceptualization, supervision, visualization, and writing (review and editing). All authors reviewed and approved the final manuscript.

\noindent\textbf{Competing interests.} The authors declare no competing interests.

\noindent\textbf{Acknowledgements.} This research was supported by a University of Pennsylvania AI Fellowship to B.L.L., research funds from the Wharton School, and cost-sharing support from the Wharton Behavioral Lab.

\section*{References}
{\footnotesize\sffamily
}

\onecolumn
\appendix
\renewcommand{\thesection}{S\arabic{section}}
\renewcommand{\thesubsection}{\thesection.\arabic{subsection}}
\setcounter{table}{0}
\renewcommand{\thetable}{S\arabic{table}}
\setcounter{figure}{0}
\renewcommand{\thefigure}{S\arabic{figure}}

\newpage
\thispagestyle{fancy}
\begin{center}
  \vspace*{1em}
  {\fontsize{12}{14}\sffamily\itshape Supplementary Information for\par}
  \vspace{6pt}
  {\fontsize{16}{18}\sffamily\bfseries
   Synthetic Contact with AI Reduces Cross-Partisan Animosity\par}
  \vspace{8pt}
  {\sffamily\normalsize
    Benjamin Lira Luttges\textsuperscript{1,\,$\ast$},\enskip
    Noah Castelo\textsuperscript{2},\enskip
    Stefano Puntoni\textsuperscript{1},\enskip
    Olivier Toubia\textsuperscript{3}\par}
  \vspace{4pt}
  {\sffamily\footnotesize
    \textsuperscript{1}The Wharton School, University of Pennsylvania.\enskip
    \textsuperscript{2}Alberta School of Business, University of Alberta.\enskip
    \textsuperscript{3}Columbia Business School, Columbia University.\par
    \vspace{2pt}
    \textsuperscript{$\ast$}Corresponding author: \href{mailto:blira@upenn.edu}{blira@upenn.edu}.\par}
\end{center}


\tableofcontents

\newpage
\section{Samples, exclusions, and deviations from preregistration}
\label{app:samples}

Each study was preregistered on AsPredicted. For each we report the registration number, the recruited and analytic samples, the preregistered exclusion rules, and any deviations from the preregistered analysis.

\paragraph*{Aversion experiment (AsPredicted~\#286{,}575).}
We preregistered a target of 500 completions; owing to a recruitment-setting error, the study was fielded for 600, yielding $N = \WYRN{}$ eligible participants. Restricting the analysis to the first 500 participants by completion time (Section~\ref{app:wyr_mlm}) leaves the estimate unchanged. The only preregistered exclusion removed participants who did not identify as Democrat or Republican, and there were no deviations from the preregistered analysis.

\paragraph*{Within-person study (AsPredicted~\#264{,}402).}
$N = \WPN{}$ partisans (\WPNDtoR{} learning about Republicans, \WPNRtoD{} learning about Democrats). The preregistration specified no exclusions from the primary analyses; all participants are retained. No deviations.

\paragraph*{Three-arm experiment (AsPredicted~\#276{,}530).}
\ExpNRecruited{} Democrats recruited; $N = \ExpN{}$ retained after the preregistered exclusions (reCAPTCHA score below $0.5$, more than two tab switches during the intervention, or completion time more than three standard deviations below the sample mean). No deviations.

\paragraph*{Behavioral experiment (AsPredicted~\#287{,}002).}
The preregistration specified a three-minute mortality-reflection comparison and a sampling frame with Republicans recruited $50/50$ with Democrats. Analyses are restricted to the preregistered sample---participants randomly assigned to the two chat conditions while preregistration~\#287{,}002 was in force ($N = \BehPreN{}$). Preregistered exclusions: reCAPTCHA below $0.5$, more than two tab switches, completion time more than three standard deviations below the mean, a bot/LLM detection flag, or sending no messages to the chatbot. No deviations.

\paragraph*{Longitudinal experiment (AsPredicted~\#288{,}592).}
Target of $1{,}200$ Democrats before exclusions; $N = \LongOneN{}$ retained at Time~1 after the preregistered exclusions (failure of any of three attention/bot checks, reCAPTCHA below $0.5$, more than two tab switches, completion time more than three standard deviations below the mean, or sending no messages), of whom $\LongTwoN{}$ returned at the one-week follow-up. Deviation: because the Wave-2 warmth distribution was floor-bunched and violated OLS assumptions, the one-week effect is reported with a rank ANCOVA (Section~\ref{app:long_rank_robust}).

\newpage
\section{Aversion Experiment}
\label{app:aversion}

\subsection{Randomization balance}
\label{app:aversion_balance}

Table~\ref{tab:aversion_balance} reports demographic characteristics by condition in the aversion experiment. No variable differed significantly across conditions ($p > .05$ for all tests), confirming successful randomization. Gender and race were not collected in this study.

\begin{table}[H]
  \centering
  \caption{Randomization balance in the aversion experiment.
           Continuous variables report $M$ ($SD$); binary variables report
           percentages. Tests are Welch's $t$-test for continuous and
           $\chi^2$ for categorical variables.}
  \label{tab:aversion_balance}
  {\small
\begin{tabular}{lcccc}
\toprule
Variable & Bot & Human & Test & $p$ \\
\midrule
Age (years) & 43.78 (13.19) & 42.76 (13.10) & $t(666)=1.01$ & $= 0.315$  \\
Democrat & 38.0\% & 44.9\% & $\chi^2(1)=2.99$ & $= 0.084$  \\
Political extremity & 73.59 (27.96) & 67.91 (31.16) & $t(441)=2.02$ & $= 0.044$  \\
Duration (min) & 9.06 (53.26) & 6.89 (11.97) & $t(369)=0.73$ & $= 0.467$  \\
\bottomrule
\end{tabular}

}
\end{table}

\subsection{Means, SDs, and correlations}
\label{app:aversion_correlations}

\begin{table}[H]
  \centering
  \caption{Means, standard deviations, and bivariate correlations among
           key variables in the aversion experiment. $^*p < .05$.}
  \label{tab:aversion_correlations}
  {\small
\begin{tabular}{lccccccc}
\toprule
Variable & $M$ & $SD$ & (1) & (2) & (3) & (4) & (5) \\
\midrule
(1) Mortality threshold (min) & 7.74 & 14.27 & --- &  &  &  &  \\
(2) Human condition (vs.\ bot) & 0.50 & 0.50 & 0.17$^*$ & --- &  &  &  \\
(3) Age & 43.27 & 13.15 & 0.01 & -0.04 & --- &  &  \\
(4) Political extremity & 70.64 & 29.77 & 0.03 & -0.10$^*$ & 0.14$^*$ & --- &  \\
(5) Duration (min) & 7.98 & 38.64 & -0.01 & -0.03 & 0.02 & -0.03 & --- \\
\bottomrule
\end{tabular}

}
\end{table}

\subsection{Primary model behind Figure~\ref{fig:wyr}}
\label{app:aversion_model}

\begin{table}[H]
  \centering
  \caption{Mortality-threshold means by condition and the regression
           \texttt{threshold $\sim$ condition} that underlies
           Figure~\ref{fig:wyr}. Coefficient is the human$-$bot
           difference in minutes.}
  \label{tab:aversion_model}
  {\small
\begin{tabular}{lcccccc}
\toprule
 & $n$ & $M$ (SD) & Coef. & 95\% CI & $t$ & $p$ \\
\midrule
Bot   & 320 & 5.27 (9.79) & ref. & --- & --- & --- \\
Human & 326 & 10.15 (17.26) & 4.88 & [2.71, 7.05] & 4.41 & $<$ .001 \\
\midrule
Cohen's $d$ &  &  & \multicolumn{4}{c}{0.35} \\
\bottomrule
\end{tabular}

}
\end{table}

\subsection{Distribution, sensitivity, and preregistered interactions}

Medians tell the same story as means but more compressed (\WYRMDhuman{} vs.\ \WYRMDbot{} minutes), indicating that the mean gap is amplified by a longer right tail among participants anticipating human contact.
A Mann--Whitney test on the raw thresholds confirms the effect non-parametrically ($p \WYRwilcoxp$).
Restricting analysis to the first \WYRfiveN{} participants by completion time leaves the estimate essentially unchanged ($\beta = \WYRfiveB$, $p \WYRfiveP$), ruling out the possibility that over-provisioning past our preregistered target of 500 drives the result.
The preregistered interaction with political extremity was not significant ($\beta = \WYRintB$, $p \WYRintP$), and the bot's acceptance advantage held within each party (Democrats: $\beta = \WYRDemB$ min, $p \WYRDemP$; Republicans: $\beta = \WYRRepB$ min, $p \WYRRepP$; no significant partner~$\times$~party interaction, $p \WYRPartyIntP$).

\subsection{Trial-level mixed-effects model}
\label{app:wyr_mlm}

The primary analysis in the main text regresses each participant's staircase-derived threshold (the mean of the last two reversals) on condition.
As a preregistered robustness check, we re-fit the model at the trial level using a mixed-effects logistic regression: $P(\text{chose chat}) = \text{logit}^{-1}(b_0 + b_1\,\text{duration} + b_2\,\text{partner} + b_3\,\text{duration}\,{\times}\,\text{partner})$ with random intercepts and slopes on duration by participant (\WYRTrialsN{} choices from \WYRN{} participants).

\paragraph{Random-effects structure.}
We include random slopes on duration (in addition to random intercepts) because participants vary in their discrimination sensitivity---the slope of each participant's psychometric curve is itself a quantity of interest in a staircase design, not a nuisance.
The data are unambiguous on this point: compared to an intercept-only specification, the random-slope model is strongly favored by a likelihood ratio test ($\chi^2(\WYRglmmLRTdf) = \WYRglmmLRTchisq$, $p \WYRglmmLRTp$; $\Delta\text{AIC} = \WYRglmmDeltaAIC$), and the intercept-only model is in fact singular (the random intercept variance collapses to zero), indicating that participant-level variability in \emph{duration sensitivity} dominates variability in baseline choice probability.

\begin{figure}[H]
  \centering
  \includegraphics[width=0.75\textwidth]{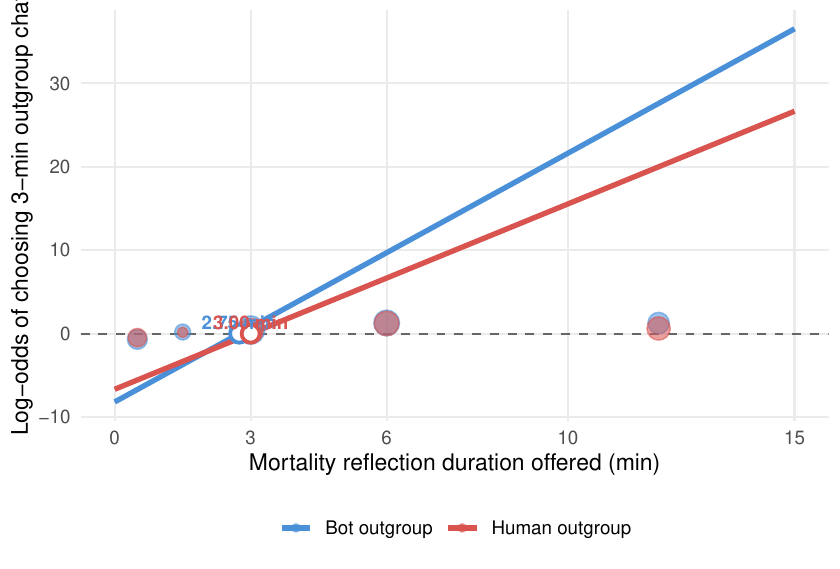}
  \caption{\textbf{Psychometric functions by condition (log-odds scale).}
           Lines: fitted linear predictor of the GLMM (random intercepts
           and slopes by participant).
           Points: observed log-odds within binned durations (size
           proportional to $n$; bins with $p \in \{0, 1\}$ dropped).
           Dashed line at $y = 0$ marks $P = 0.5$; white circles mark
           each curve's crossing.}
  \label{fig:wyr_psych}
\end{figure}

\paragraph{Results.}
Figure~\ref{fig:wyr_psych} shows the fitted linear predictor (log-odds of choosing the three-minute chat) as a function of offered mortality duration, by condition.
The two lines cross zero at the psychometric midpoint: bot participants are indifferent at $X = \WYRglmmThresholdBot{}$ min; human participants at $X = \WYRglmmThresholdHuman{}$ min.
The slope of the human condition is shallower than that of the bot condition (duration~$\times$~partner interaction: $\beta = \WYRglmmInterB$, $p \WYRglmmInterP$), which is consistent with the main-text finding: the psychometric curve rises more slowly in the human condition, so at any given mortality duration above the indifference point, a smaller share of human-condition participants have switched over to choosing chat.
The psychometric midpoints are close across conditions because the threshold distribution is heavily right-skewed (medians \WYRMDbot{} and \WYRMDhuman{} min, against means of \WYRMbot{} and \WYRMhuman{}): the typical participant is only modestly more human-averse, while the condition difference is carried by the shallower human slope and a heavier right tail of strongly human-averse participants---which the mean threshold in the main text reflects, and a single midpoint cannot.

\subsection{Sample cross-party conversations}
\label{app:wyr_transcripts}

In the human condition, participants who reached the chat task were paired with a real outgroup partner for a three-minute live text conversation about immigration.
Two verbatim exchanges appear below to illustrate the range of dialogue these participants produced.
The first shows policy-level disagreement bounded by mutual civility; the second shows partial within-party dissent that surfaces in cross-party contact.

\begin{quote}\small
\textbf{Democrat:} \emph{Hi there, how are you?}\\[2pt] \textbf{Republican:} \emph{Hi! I am good how are you}\\[2pt] \textbf{Democrat:} \emph{Doing well, so what is your take on immigration?}\\[2pt] \textbf{Democrat:} \emph{I personally think we need to cut down on illegal immigration but do it in a far more humane way than Trump and the Republicans have been handling it}\\[2pt] \textbf{Republican:} \emph{I think illegal immigration has to be curbed almost completely, and legal immigration should be encouraged but largely based on skill}\\[2pt] \textbf{Democrat:} \emph{We should not have people coming across the border illegally, I agree 100\%}\\[2pt] \textbf{Republican:} \emph{I do not think adding a ton of unskilled workers or non-working people is a general positive for the country, so I think we should have like a skill based immigration system}\\[2pt] \textbf{Democrat:} \emph{But sending in violent ICE agents to get rid of them does nobody any good}\\[2pt] \textbf{Republican:} \emph{That being said I do disagree with the Trump admin on giving out a ton of H1B visas}\\[2pt] \textbf{Republican:} \emph{How would you suggest we get rid of them?}\\[2pt] \textbf{Democrat:} \emph{Send in local police forces that actually follow the law}\\[2pt] \textbf{Republican:} \emph{I see. It would require a lot of coordination between the federal and local governments and I am not sure they could force local police to do it because of separation of state and federal, but if they could I would support it}
\end{quote}

\begin{quote}\small
\textbf{Democrat:} \emph{Hi, whats your thoughts on immigration}\\[2pt] \textbf{Republican:} \emph{Hello.}\\[2pt] \textbf{Democrat:} \emph{I think we need to re do our immigration system}\\[2pt] \textbf{Republican:} \emph{I think that immigration is good for our country. I do disagree with illegal immigration however.}\\[2pt] \textbf{Republican:} \emph{What are your thoughts?}\\[2pt] \textbf{Democrat:} \emph{I agree. I think people should come here the right way. I just don't think its as easy as people think it is}\\[2pt] \textbf{Republican:} \emph{I also agree with that. I know that isn't very popular among other `Republicans', which is why I cannot fully identify as a Republican.}\\[2pt] \textbf{Democrat:} \emph{I feel the same way about democrats. I do agree with most other policies}\\[2pt] \textbf{Republican:} \emph{I think that we should help with barriers to immigration and integrate immigrants so that they can be here the right way and have proper protections and resources.}
\end{quote}

Across the cross-party conversations, no exchange contained personal incivility, profanity, or name-calling directed at the conversation partner: people disagreed substantively but treated each other politely.
For sample outgroup-bot conversations of the kind used in the bot condition, see Appendix~\ref{app:disconfirm_excerpts}.

\section{Within-Person Study}
\label{app:within_person}

\subsection{Sample descriptives}
\label{app:descriptives}

Table~\ref{tab:descriptives} reports demographic characteristics and pre-interaction outgroup warmth by learner party.

\begin{table}[H]
  \centering
  \caption{Sample characteristics by learner party.
           Values in parentheses are standard deviations.}
  \label{tab:descriptives}
  
\begin{tblr}[         
]                     
{                     
colspec={Q[]Q[]Q[]},
hline{2}={1-3}{solid, black, 0.05em},
hline{1}={1-3}{solid, black, 0.08em},
hline{8}={1-3}{solid, black, 0.08em},
}                     
& Democrats & Republicans \\
$N$ & 248 & 252 \\
Age & 43.3 (14.0) & 43.8 (13.1) \\
Female (\%) & 53 & 52 \\
White (\%) & 71 & 81 \\
Extremity (0--100) & 40.0 (13.0) & 36.0 (14.9) \\
Pre-warmth (0--100) & 22.6 (22.8) & 35.9 (24.4) \\
\end{tblr}

\end{table}

\subsection{Means and misperception tests behind Figure~\ref{fig:wp_main}A}
\label{app:misperception_means}

\begin{table}[H]
  \centering
  \caption{Own-party attitudes and outgroup-perceived attitudes on the
           6-item GREEN environmental scale (1--5). Each party's
           outgroup belief is compared against the outgroup's own
           reported attitude via Welch's $t$-test, the same comparison
           visualized in Figure~\ref{fig:wp_main}A.}
  \label{tab:misperception_means}
  {\small
\begin{tabular}{lcccc}
\toprule
Quantity & $M$ (SD) & $n$ & $t$ vs.\ reality & $p$ \\
\midrule
Republican attitudes (own report)        & 3.73 (1.06) & 252 & --- & --- \\
Democrats' estimate of Republicans       & 2.16 (1.03) & 248 & -16.79 & $<$ .001 \\
\addlinespace[4pt]
Democratic attitudes (own report)        & 4.15 (0.77) & 248 & --- & --- \\
Republicans' estimate of Democrats       & 3.96 (0.93) & 252 & -2.46 & 0.014 \\
\bottomrule
\end{tabular}

}
\end{table}

\subsection{Correlations}
\label{app:correlations}

Table~\ref{tab:correlations} reports means, standard deviations, and bivariate correlations among the key study variables.

\begin{table}[H]
  \centering
  \caption{Means, standard deviations, and bivariate correlations among study variables.
           $^*p < .05$.}
  \label{tab:correlations}
  {\small
\begin{tblr}[         
]                     
{                     
colspec={Q[]Q[]Q[]Q[]Q[]Q[]Q[]Q[]Q[]Q[]Q[]Q[]},
hline{2}={1-12}{solid, black, 0.05em},
hline{1}={1-12}{solid, black, 0.08em},
hline{11}={1-12}{solid, black, 0.08em},
}                     
& $M$ & $SD$ & 1 & 2 & 3 & 4 & 5 & 6 & 7 & 8 & 9 \\
1. Pre-warmth & 29.32 & 24.54 & --- &  &  &  &  &  &  &  &  \\
2. Post-warmth & 33.66 & 25.70 & 0.89* & --- &  &  &  &  &  &  &  \\
3. Pre-accuracy & -1.19 & 0.89 & 0.49* & 0.45* & --- &  &  &  &  &  &  \\
4. Post-accuracy & -0.80 & 0.72 & 0.32* & 0.35* & 0.46* & --- &  &  &  &  &  \\
5. Extremity & 38.02 & 14.14 & -0.31* & -0.27* & -0.20* & -0.16* & --- &  &  &  &  \\
6. Informativeness & 3.84 & 1.06 & 0.22* & 0.29* & 0.22* & 0.25* & 0.03 & --- &  &  &  \\
7. Empathy & 3.53 & 1.04 & 0.13* & 0.18* & 0.04 & 0.13* & -0.01 & 0.35* & --- &  &  \\
8. Bot turns & 11.11 & 6.00 & -0.03 & -0.05 & -0.05 & -0.04 & -0.04 & -0.08 & -0.05 & --- &  \\
9. Words written & 199.86 & 123.31 & -0.02 & -0.04 & -0.03 & -0.02 & 0.02 & -0.03 & 0.06 & 0.18* & --- \\
\end{tblr}
}
\end{table}

\subsection{Mixed-effects models}
\label{app:main_models}

\begin{table}[H]
  \centering
  \caption{Mixed-effects models predicting outgroup warmth and belief accuracy.
           Belief accuracy is the negative absolute error between a participant's
           guess of the outgroup's environmental attitudes and the outgroup's
           actual mean (higher = more accurate).
           Columns~(1) and~(4) include only the pre-to-post time effect.
           Columns~(2) and~(5) add the Learner party interaction.
           Columns~(3) and~(6) additionally include political extremity
           (centered) and its interaction with time.
           Standard errors in parentheses.
           Reference group: Democrats learning about Republicans, pre-interaction.}
  \label{tab:main}
  
\begin{talltblr}[         
entry=none,label=none,
note{}={* p $<$ 0.05, ** p $<$ 0.01, *** p $<$ 0.001},
]                     
{                     
colspec={Q[]Q[]Q[]Q[]Q[]Q[]Q[]},
hline{2}={3,6-7}{solid, black, 0.03em},
hline{2}={2,5}{solid, black, 0.03em, l=-0.5},
hline{2}={4}{solid, black, 0.03em, r=-0.5},
hline{3}={1-7}{solid, black, 0.05em},
hline{15}={1-7}{solid, black, 0.05em},
hline{1}={1-7}{solid, black, 0.08em},
hline{16}={1-7}{solid, black, 0.08em},
column{3-4,6-7}={}{halign=c},
cell{1}{1}={}{halign=c},
cell{1}{2}={c=3}{halign=c},
cell{1}{5}={c=3}{halign=c},
cell{2-15}{1}={}{halign=l},
cell{2-15}{2}={}{halign=c},
cell{2-15}{5}={}{halign=c},
}                     
& Outgroup warmth &  &  & Belief accuracy &  &  \\
& (1) & (2) & (3) & (4) & (5) & (6) \\
Intercept & 29.316*** & 22.601*** & 23.574*** & -1.189*** & -1.680*** & -1.665*** \\
& (1.123) & (1.543) & (1.497) & (0.036) & (0.045) & (0.045) \\
Post (vs.\ pre) & 4.342*** & 5.149*** & 5.055*** & 0.393*** & 0.628*** & 0.624*** \\
& (0.524) & (0.743) & (0.745) & (0.038) & (0.052) & (0.052) \\
R$\to$D (vs.\ D$\to$R) &  & 13.324*** & 11.393*** &  & 0.976*** & 0.946*** \\
&  & (2.174) & (2.119) &  & (0.063) & (0.064) \\
Post $\times$ R$\to$D &  & -1.602 & -1.415 &  & -0.465*** & -0.458*** \\
&  & (1.046) & (1.055) &  & (0.073) & (0.073) \\
Extremity (centered) &  &  & -0.480*** &  &  & -0.007*** \\
&  &  & (0.075) &  &  & (0.002) \\
Post $\times$ Extremity &  &  & 0.046 &  &  & 0.002 \\
&  &  & (0.037) &  &  & (0.003) \\
Num.Obs. & 1000 & 1000 & 1000 & 1000 & 1000 & 1000 \\
\end{talltblr}

\end{table}

\subsection{Ingroup warmth and affective polarization}
\label{app:ingroup}

Our primary measure is warmth toward the outgroup, which a shift in feelings toward one's own party cannot mechanically inflate. For completeness, we nonetheless confirm that the gain reflects genuine outgroup warming rather than a general flattening of partisan affect, and that it also holds on the standard affective-polarization index (ingroup minus outgroup warmth, where higher values indicate more polarization).

Ingroup warmth decreased only slightly after the chatbot interaction ($b = \IngroupTimeB{}$, $SE = \IngroupTimeSE{}$, $p \IngroupTimeP{}$)---a fraction of the \WPWarmthB{}-point rise in outgroup warmth---and the change did not differ between Democrats and Republicans ($b = \IngroupIntB{}$, $SE = \IngroupIntSE{}$, $p \IngroupIntP{}$).
Affective polarization declined in both groups: among Democrats, the index fell from \APPreMDtoR{} to \APPostMDtoR{} (a change of \APChangeMDtoR{}); among Republicans, from \APPreMRtoD{} to \APPostMRtoD{} (a change of \APChangeMRtoD{}).
A mixed-effects model confirmed the decline was significant ($b = \APTimeB{}$, $SE = \APTimeSE{}$, $p \APTimeP{}$), and both groups depolarized at similar rates ($b = \APIntB{}$, $SE = \APIntSE{}$, $p \APIntP{}$; Table~\ref{tab:appendix_ingroup}).

The warmth gains reported in the main analysis reflect genuine improvement toward the outgroup, not compression of the feeling thermometer.

\begin{table}[H]
  \centering
  \caption{Mixed-effects models predicting ingroup warmth and the affective
           polarization index (ingroup minus outgroup warmth) before and after
           chatbot interaction.
           Standard errors in parentheses.
           Reference group: Democrats learning about Republicans, pre-interaction.}
  \label{tab:appendix_ingroup}
  
\begin{talltblr}[         
entry=none,label=none,
note{}={* p $<$ 0.05, ** p $<$ 0.01, *** p $<$ 0.001},
]                     
{                     
colspec={Q[]Q[]Q[]},
hline{2}={1-3}{solid, black, 0.05em},
hline{10}={1-3}{solid, black, 0.05em},
hline{1}={1-3}{solid, black, 0.08em},
hline{11}={1-3}{solid, black, 0.08em},
column{2-3}={}{halign=c},
column{1}={}{halign=l},
}                     
& Ingroup warmth & Affective polarization \\
Intercept & 76.016*** & 53.415*** \\
& (1.266) & (1.979) \\
Post (vs.\ pre) & -1.069* & -6.218*** \\
& (0.540) & (0.941) \\
R$\to$D (vs.\ D$\to$R) & 4.988** & -8.336** \\
& (1.783) & (2.787) \\
Post $\times$ R$\to$D & -0.590 & 1.011 \\
& (0.760) & (1.326) \\
Num.Obs. & 1000 & 1000 \\
\end{talltblr}

\end{table}

\subsection{Individual heterogeneity}
\label{app:heterogeneity}

\begin{figure}[H]
  \centering
  \includegraphics[width=\linewidth]{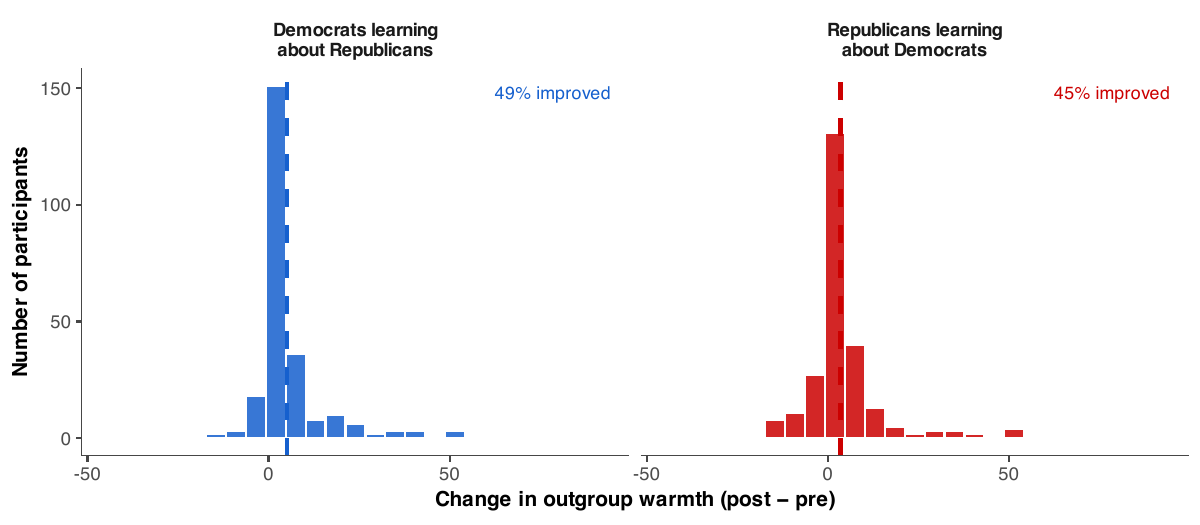}
  \caption{Distribution of individual-level change in outgroup warmth (post $-$ pre),
           by learner party.
           Dashed lines mark the group mean.
           Annotations show the percentage of participants with a positive change.}
  \label{fig:hist_warmth}
\end{figure}

The mean warmth gain masks variation across individuals.
Figure~\ref{fig:hist_warmth} plots the distribution of individual-level warmth change (post minus pre) for each group. Among Democrats, \PctImprovedDtoR\% showed a positive change in outgroup warmth; among Republicans, \PctImprovedRtoD\% did.
About half of each group warmed toward the outgroup; most of the remainder were unchanged, and only a minority grew colder.
This distribution confirms that the mean effect is not driven by a handful of outliers.

\subsection{Bot--item semantic proximity and belief updating}
\label{app:content_dose}

A natural worry about the conversation-content claim is that we have only inferred topic exposure from end-of-chat ratings of the bot. We address this here by measuring topic exposure directly from the transcripts and testing whether conversations that hewed closer to the six GREEN items produced more belief updating and more warmth gain. We embed each of the six GREEN items and each of the \WPProxNTurns{} bot turns across the \WPProxN{} conversations using OpenAI's \texttt{text-embedding-3-large}. For every participant we compute a single proximity score: for each item, the maximum cosine similarity between that item and any of the participant's bot turns, then averaged across the six items (i.e., ``did the bot hit every item at some point''):
\begin{equation}
\text{proximity}_p = \frac{1}{6} \sum_{i=1}^{6} \max_{t \in T_p} \cos\!\left(\mathbf{e}_i, \mathbf{e}_t\right),
\end{equation}
where $\mathbf{e}_i$ is the embedding of GREEN item $i$, $T_p$ is the set of participant $p$'s bot turns, and $\mathbf{e}_t$ is the embedding of turn $t$.

Pooled across both directions, proximity is associated with belief accuracy gain ($r = \WPProxAccRPool{}$, $p \WPProxAccPPool{}$) but not warmth gain ($r = \WPProxWarmRPool{}$, $p \WPProxWarmPPool{}$; Table~\ref{tab:proximity}, Panel A). The accuracy effect is concentrated in the D$\to$R direction ($r = \WPProxAccRDtoR{}$, $p \WPProxAccPDtoR{}$); in R$\to$D it is null ($r = \WPProxAccRRtoD{}$, $p \WPProxAccPRtoD{}$), and the party $\times$ proximity interaction is significant ($b = \WPProxIntAccB{}$, $SE = \WPProxIntAccSE{}$, $p \WPProxIntAccP{}$). Belief updating and warmth gain themselves are coupled only for Democrats ($r = \WPAccWarmRDtoR{}$, $p \WPAccWarmPDtoR{}$); for Republicans they are independent ($r = \WPAccWarmRRtoD{}$, $p \WPAccWarmPRtoD{}$).

A second, independent content measure points the same way. Coding every bot message for whether its substance contradicted (\emph{disconfirming}), reinforced (\emph{confirming}), or was irrelevant to (\emph{neutral}) the outgroup stereotype, we find confirming content was negligible---only 1.5\% of bot messages---so conversations varied chiefly in \emph{how much} stereotype-disconfirming substance the bot delivered, not in whether it also confirmed. Among Democrats learning about Republicans, conversations carrying more disconfirming substance produced larger accuracy gains (standardized $b = 0.16$, $p = .003$, controlling baseline accuracy; the per-message disconfirming proportion trends in the same direction, $p = .11$); in R$\to$D the association is null. As with proximity, this measure is correlational and post-randomization---more disconfirming content could be both a cause and a marker of an engaged participant---so it corroborates rather than proves the content route.

We cannot test whether disconfirming content mediates the \emph{between-condition} warmth effect directly: control-bot conversations carry virtually no stereotype-disconfirming substance ($0.1\%$ of cats-and-dogs messages vs.\ $\DisconfirmPctDtoR\%$ of outgroup-bot messages), so the candidate mediator is, by design, collinear with condition and the model is not identified. Within the outgroup-bot conversations we can instead ask whether variation in disconfirming content tracks warmth gain through belief updating. Among Democrats learning about Republicans ($N = \MedN{}$), conversations carrying more disconfirming content produced marginally larger accuracy gains ($a = \MedDoseAccB{}$, $p \MedDoseAccP{}$), accuracy gains strongly predicted warmth gains ($b = \MedAccWarmthB{}$, $p \MedAccWarmthP{}$), and the bootstrapped indirect path was positive but not significant ($a\cdot b = \MedIndirectB{}$, 95\% CI $[\MedIndirectLo{}, \MedIndirectHi{}]$, $p \MedIndirectP{}$). The data are thus consistent with a content$\to$belief$\to$warmth route but underpowered to confirm it at the conversation level. In the R$\to$D direction the same model is null at every path ($N = \MedNRtoD{}$; indirect $a\cdot b = \MedIndirectBRtoD{}$, 95\% CI $[\MedIndirectLoRtoD{}, \MedIndirectHiRtoD{}]$)---as expected, since Republicans begin nearly accurate about Democrats and their warmth gains do not track belief updating in the first place.

\begin{figure}[H]
  \centering
  \includegraphics[width=0.85\linewidth]{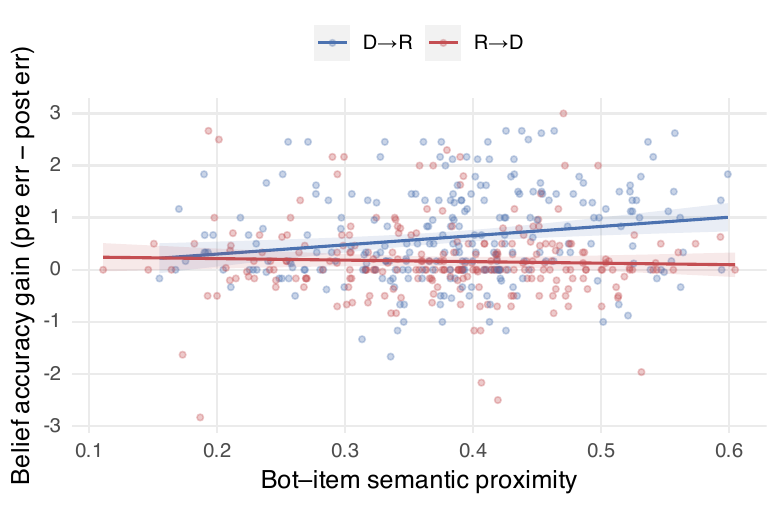}
  \caption{Conversation-level bot--item semantic proximity vs. belief
           accuracy gain, by learner direction. Points are individual
           participants; lines are OLS fits with 95\% confidence bands.
           Proximity tracks accuracy gain in D$\to$R but not
           in R$\to$D.}
  \label{fig:proximity_accuracy}
\end{figure}

In the D$\to$R sample, belief updating accounts for the link between conversation content and warmth (Figure~\ref{fig:proximity_accuracy}, Table~\ref{tab:proximity}, Panel B). Conversations with more on-topic bot turns produce larger accuracy gains ($a = \WPMedA{}$, $SE = \WPMedASE{}$, $p \WPMedAP{}$); accuracy gain in turn predicts warmth gain controlling for proximity ($b = \WPMedB{}$, $SE = \WPMedBSE{}$, $p \WPMedBP{}$); proximity has no direct effect on warmth once accuracy is held constant ($c' = \WPMedCp{}$, $p \WPMedCpP{}$). The bootstrapped indirect path is significant ($a \cdot b = \WPMedInd{}$, 95\% CI $[\WPMedIndLow{}, \WPMedIndHigh{}]$, $p \WPMedIndP{}$), and the total effect is not, consistent with full indirect-only mediation. In the R$\to$D sample all three paths are null. The asymmetry mirrors the by-party gradient in accuracy gain reported in Section~\ref{app:misperception_means}: where belief updating happens, content-driven proximity tracks it, and warmth tracks updating in turn.

\begin{table}[H]
  \centering
  \caption{Bot--item semantic proximity, belief updating, and warmth gain.
           Panel A reports Pearson correlations between conversation-level
           proximity and the two outcomes, pooled and by learner direction.
           Panel B reports the mediation model proximity~$\to$~accuracy
           gain~$\to$~warmth gain in both learner directions, with bootstrapped
           (2{,}000 draws) confidence intervals for the indirect effect.
           All variables standardized.}
  \label{tab:proximity}
  {\small
\begin{tblr}{
  colspec = {l l r r r},
  row{1,8} = {font=\bfseries},
  row{1,8} = {bg=gray!10},
  hline{2} = {1pt},
  hline{8} = {1pt},
  hline{9} = {0.5pt},
  hline{15} = {0.5pt},
}
\textbf{Panel A: Correlations} & & N & $r$ & $p$ \\
Pooled & Accuracy gain & 500 & 0.09 & 0.043 \\
       & Warmth gain   & 500 & 0.03 & 0.525 \\
D$\to$R & Accuracy gain & 248 & 0.19 & 0.003 \\
       & Warmth gain   & 248 & 0.04 & 0.571 \\
R$\to$D & Accuracy gain & 252 & -0.04 & 0.546 \\
       & Warmth gain   & 252 & 0.02 & 0.787 \\
\textbf{Panel B: Mediation} & Est. & SE & 95\% CI & $p$ \\
\emph{Democrats $\to$ Republicans} & & & & \\
\quad Proximity $\to$ Accuracy gain ($a$) & 0.19 & 0.06 & [0.08, 0.31] & < .001 \\
\quad Accuracy gain $\to$ Warmth gain ($b$) & 0.22 & 0.08 & [0.06, 0.37] & 0.007 \\
\quad Proximity $\to$ Warmth gain (direct, $c'$) & -0.00 & 0.06 & [-0.13, 0.12] & 0.940 \\
\quad Indirect $a\cdot b$ & 0.04 & 0.02 & [0.01, 0.09] & 0.037 \\
\quad Total & 0.04 & 0.06 & [-0.07, 0.15] & 0.525 \\
\emph{Republicans $\to$ Democrats} & & & & \\
\quad Proximity $\to$ Accuracy gain ($a$) & -0.04 & 0.07 & [-0.19, 0.10] & 0.594 \\
\quad Accuracy gain $\to$ Warmth gain ($b$) & -0.01 & 0.06 & [-0.13, 0.10] & 0.890 \\
\quad Proximity $\to$ Warmth gain (direct, $c'$) & 0.02 & 0.06 & [-0.11, 0.14] & 0.791 \\
\quad Indirect $a\cdot b$ & 0.00 & 0.00 & [-0.01, 0.01] & 0.947 \\
\quad Total & 0.02 & 0.06 & [-0.11, 0.14] & 0.783 \\
\end{tblr}
}
\end{table}

\section{Three-Arm Experiment}
\label{app:experiment}

\subsection{Randomization balance}
\label{app:exp_balance}

Table~\ref{tab:exp_balance} reports demographic characteristics by condition.
No variable differed significantly across conditions ($p > .05$ for all tests), confirming successful randomization.

\begin{table}[H]
  \centering
  \caption{Randomization balance across experimental conditions.
           Continuous variables report $M$ ($SD$); binary variables report
           percentages. Tests are one-way ANOVA ($F$) for continuous and
           $\chi^2$ for categorical variables.}
  \label{tab:exp_balance}
  {\small
\begin{tblr}[         
]                     
{                     
colspec={Q[]Q[]Q[]Q[]Q[]Q[]},
hline{2}={1-6}{solid, black, 0.05em},
hline{1}={1-6}{solid, black, 0.08em},
hline{7}={1-6}{solid, black, 0.08em},
}                     
& Outgroup bot & Cats/dogs bot & Space Invaders & Statistic & $p$ \\
Age & 40.5 (14.4) & 39.8 (13.5) & 40.3 (14.4) & 0.14 & .873 \\
Female (\%) & 62 & 64 & 65 & 0.33 & .847 \\
White (\%) & 75 & 71 & 69 & 1.89 & .389 \\
Extremity (0--100) & 82.1 (20.5) & 79.8 (21.7) & 77.2 (24.5) & 2.56 & .078 \\
Duration (sec) & 764.9 (232.4) & 761.9 (249.2) & 742.4 (184.2) & 0.68 & .506 \\
\end{tblr}
}
\end{table}

\subsection{Correlations}
\label{app:exp_correlations}

Table~\ref{tab:exp_correlations} reports bivariate correlations among key variables in the three-arm experiment.

\begin{table}[H]
  \centering
  \caption{Bivariate correlations among key variables in the three-arm
           experiment. Upper triangle reports Pearson $r$.
           $^*p < .05$, $^{**}p < .01$, $^{***}p < .001$.}
  \label{tab:exp_correlations}
  {\small
\begin{tblr}[         
]                     
{                     
colspec={Q[]Q[]Q[]Q[]Q[]Q[]Q[]Q[]Q[]Q[]Q[]},
hline{2}={1-11}{solid, black, 0.05em},
hline{1}={1-11}{solid, black, 0.08em},
hline{10}={1-11}{solid, black, 0.08em},
}                     
& $M$ & $SD$ & 1 & 2 & 3 & 4 & 5 & 6 & 7 & 8 \\
1. Outgroup warmth & 20.57 & 20.85 & --- &  &  &  &  &  &  &  \\
2. Ingroup warmth & 76.43 & 19.02 & 0.01 & --- &  &  &  &  &  &  \\
3. Age & 40.16 & 14.06 & -0.03 & 0.19* & --- &  &  &  &  &  \\
4. Extremity & 79.60 & 22.43 & -0.13* & 0.58* & 0.17* & --- &  &  &  &  \\
5. Duration (sec) & 755.97 & 223.28 & 0.05 & 0.09* & 0.04 & 0.03 & --- &  &  &  \\
6. Bot: stereotype-disconfirming & 2.21 & 1.44 & 0.20* & 0.09 & 0.04 & 0.11* & -0.06 & --- &  &  \\
7. Bot: empathy & 3.88 & 0.37 & -0.07 & 0.02 & 0.06 & 0.02 & -0.05 & -0.04 & --- &  \\
8. Bot: friendliness & 4.41 & 0.50 & 0.02 & 0.14* & -0.04 & 0.05 & -0.01 & -0.01 & 0.12* & --- \\
\end{tblr}
}
\end{table}

\subsection{Main-effects regression}
\label{app:exp_main_table}

\begin{table}[H]
  \centering
  \caption{OLS models predicting outgroup warmth from condition assignment.
           Model~(1) uses the outgroup bot as reference; Model~(2) uses the
           cats-and-dogs control. Standard errors in parentheses.}
  \label{tab:exp_main}
  
\begin{talltblr}[         
entry=none,label=none,
note{}={* p $<$ 0.05, ** p $<$ 0.01, *** p $<$ 0.001},
]                     
{                     
colspec={Q[]Q[]Q[]},
hline{2}={1-3}{solid, black, 0.05em},
hline{10}={1-3}{solid, black, 0.05em},
hline{1}={1-3}{solid, black, 0.08em},
hline{13}={1-3}{solid, black, 0.08em},
column{2-3}={}{halign=c},
column{1}={}{halign=l},
}                     
& Outgroup bot ref. & Cats/dogs ref. \\
Intercept & 29.187*** & 16.967*** \\
& (1.430) & (1.289) \\
Cats/dogs bot & -12.220*** &  \\
& (1.925) &  \\
Space Invaders & -12.098*** & 0.121 \\
& (1.938) & (1.836) \\
Outgroup bot &  & 12.220*** \\
&  & (1.925) \\
Num.Obs. & 679 & 679 \\
R2 & 0.070 & 0.070 \\
R2 Adj. & 0.068 & 0.068 \\
\end{talltblr}

\end{table}

\subsection{Verbatim examples of disconfirming content}
\label{app:disconfirm_excerpts}

Three verbatim excerpts from Republican-bot conversations with Democratic learners in the three-arm experiment, drawn from messages coded as stereotype-disconfirming.

\begin{quote}\small
\textbf{User:} \emph{I see traditional values as centered around family and religion, as being closed minded and putting others down.}\\[2pt] \textbf{Republican bot:} \emph{Family and faith have always been core values for a lot of people in the Republican Party\ldots{} However, I think it's important to recognize that there are many different interpretations of those values. Not everyone in the party is close-minded or dismissive of others; many people are working hard to find common ground and build connections.}
\end{quote}

\begin{quote}\small
\textbf{User:} \emph{I'm hoping we could discuss your opinion regarding environmental policy.}\\[2pt] \textbf{Republican bot:} \emph{I think it's important to protect the environment, but I also believe in balancing that with economic growth and job creation\ldots{} practical solutions that encourage innovation, like promoting clean energy technologies without going overboard with regulations.}
\end{quote}

\begin{quote}\small
\textbf{User:} \emph{The current republican party has a ton of control and republicans are just letting it happen.}\\[2pt] \textbf{Republican bot:} \emph{There's definitely a lot of discussion around how the Republican Party has changed\ldots{} Some folks feel that the party has shifted too far in one direction or that it's not representing their values as well as it used to.}
\end{quote}

\section{Behavioral Experiment}
\label{app:behavioral}

\subsection{Randomization balance}
\label{app:beh_balance}

Table~\ref{tab:beh_balance} reports demographic characteristics by condition in the behavioral experiment. No variable differed significantly across conditions ($p > .05$ for all tests), confirming successful randomization.

\begin{table}[H]
  \centering
  \caption{Randomization balance in the behavioral experiment.
           Continuous variables report $M$ ($SD$); binary variables report
           percentages. Tests are one-way ANOVA ($F$) for continuous and
           $\chi^2$ for categorical variables.}
  \label{tab:beh_balance}
  {\small
\begin{tabular}{lcccc}
\toprule
Variable & Outgroup bot & Cats/dogs & Test & $p$ \\
\midrule
Age (years) & 38.82 (14.20) & 38.28 (13.25) & $F(1,1713)=0.67$ & $= 0.415$  \\
Female & 61.9\% & 62.7\% & $\chi^2(1)=0.08$ & $= 0.774$  \\
White & 70.7\% & 74.8\% & $\chi^2(1)=3.41$ & $= 0.065$  \\
Political extremity & 78.19 (17.82) & 77.79 (18.58) & $F(1,1752)=0.22$ & $= 0.641$  \\
Duration (min) & 26.24 (460.56) & 10.42 (4.55) & $F(1,1752)=1.03$ & $= 0.310$  \\
\bottomrule
\end{tabular}

}
\end{table}

\subsection{Means, SDs, and correlations}
\label{app:beh_correlations}

\begin{table}[H]
  \centering
  \caption{Means, standard deviations, and bivariate correlations among
           key variables in the behavioral experiment, including the four
           coded bot-content dimensions. $^*p < .05$.}
  \label{tab:beh_correlations}
  {\footnotesize
\begin{tabular}{lcccccccccc}
\toprule
Variable & $M$ & $SD$ & (1) & (2) & (3) & (4) & (5) & (6) & (7) & (8) \\
\midrule
(1) Chose outgroup conversation & 0.66 & 0.47 & --- &  &  &  &  &  &  &  \\
(2) Outgroup bot (vs.\ cats/dogs) & 0.50 & 0.50 & 0.10$^*$ & --- &  &  &  &  &  &  \\
(3) Age & 38.56 & 13.74 & -0.01 & 0.02 & --- &  &  &  &  &  \\
(4) Party (Rep $\uparrow$) & 76.59 & 17.40 & 0.04 & 0.01 & 0.12$^*$ & --- &  &  &  &  \\
(5) Political extremity & 78.06 & 18.17 & 0.02 & 0.01 & 0.09$^*$ & 1.00$^*$ & --- &  &  &  \\
(6) Bot: stereotype-disconfirming & 2.04 & 1.38 & 0.07$^*$ & 0.51$^*$ & 0.00 & 0.01 & 0.05$^*$ & --- &  &  \\
(7) Bot: empathy & 3.85 & 0.66 & 0.04 & 0.38$^*$ & -0.05$^*$ & 0.02 & 0.03 & 0.40$^*$ & --- &  \\
(8) Bot: friendliness & 4.17 & 0.55 & 0.07$^*$ & 0.42$^*$ & -0.02 & -0.02 & 0.00 & 0.25$^*$ & 0.38$^*$ & --- \\
\bottomrule
\end{tabular}

}
\end{table}

\subsection{Primary model behind Figure~\ref{fig:behavioral_mt3}}
\label{app:behavioral_model}

\begin{table}[H]
  \centering
  \caption{Preregistered logistic regression
           \texttt{outgroup\_chat $\sim$ condition + party + extremity\_c}
           that underlies Figure~\ref{fig:behavioral_mt3}.}
  \label{tab:behavioral_model}
  {\small
\begin{tabular}{lcccc}
\toprule
Term & Coef. (logit) & OR & 95\% CI (OR) & $p$ \\
\midrule
Intercept & 0.40 & 1.49 & [1.21, 1.83] & $< .001$ \\
Outgroup bot (vs.\ cats/dogs) & 0.29 & 1.33 & [1.04, 1.71] & $= 0.025$ \\
Republican (vs.\ Democrat) & 0.07 & 1.07 & [0.83, 1.37] & $= 0.610$ \\
Political extremity (centered) & 0.00 & 1.00 & [0.99, 1.01] & $= 0.805$ \\
\midrule
$N$ & \multicolumn{4}{c}{1069} \\
\bottomrule
\end{tabular}
}
\end{table}

\section{Longitudinal Experiment}
\label{app:longitudinal}

\subsection{Randomization balance}
\label{app:long_balance}

Table~\ref{tab:long_balance} reports demographic characteristics by condition in the longitudinal experiment. No variable differed significantly across conditions ($p > .05$ for all tests), confirming successful randomization.

\begin{table}[H]
  \centering
  \caption{Randomization balance in the longitudinal experiment.
           Continuous variables report $M$ ($SD$); binary variables report
           percentages. Tests are one-way ANOVA ($F$) for continuous and
           $\chi^2$ for categorical variables.}
  \label{tab:long_balance}
  {\small
\begin{tabular}{lcccc}
\toprule
Variable & Outgroup bot & Cats/dogs & Test & $p$ \\
\midrule
Age (years) & 40.18 (12.95) & 39.63 (13.40) & $F(1,1194)=0.52$ & $= 0.472$  \\
Female & 62.1\% & 60.7\% & $\chi^2(1)=0.20$ & $= 0.654$  \\
White & 65.0\% & 68.3\% & $\chi^2(1)=1.45$ & $= 0.229$  \\
Political extremity & 80.33 (17.83) & 80.07 (18.35) & $F(1,1280)=0.07$ & $= 0.794$  \\
Duration (min) & 21.01 (278.48) & 7.75 (4.78) & $F(1,1280)=1.45$ & $= 0.229$  \\
\bottomrule
\end{tabular}

}
\end{table}

\subsection{Means, SDs, and correlations}
\label{app:long_correlations}

\begin{table}[H]
  \centering
  \caption{Means, standard deviations, and bivariate correlations among
           key variables in the longitudinal experiment, including the
           four coded bot-content dimensions. $^*p < .05$.}
  \label{tab:long_correlations}
  {\footnotesize
\begin{tabular}{lcccccccccc}
\toprule
Variable & $M$ & $SD$ & (1) & (2) & (3) & (4) & (5) & (6) & (7) & (8) \\
\midrule
(1) Outgroup warmth (post-chat) & 24.64 & 23.89 & --- &  &  &  &  &  &  &  \\
(2) Outgroup warmth (baseline) & 21.09 & 22.83 & 0.71$^*$ & --- &  &  &  &  &  &  \\
(3) Outgroup bot (vs.\ cats/dogs) & 0.50 & 0.50 & 0.23$^*$ & 0.04 & --- &  &  &  &  &  \\
(4) Age & 39.90 & 13.18 & -0.02 & -0.07$^*$ & 0.02 & --- &  &  &  &  \\
(5) Political extremity & 80.10 & 18.14 & -0.15$^*$ & -0.18$^*$ & 0.01 & 0.17$^*$ & --- &  &  &  \\
(6) Bot: stereotype-disconfirming & 2.07 & 1.39 & 0.07$^*$ & -0.06$^*$ & 0.51$^*$ & 0.03 & 0.08$^*$ & --- &  &  \\
(7) Bot: empathy & 3.75 & 0.69 & 0.05 & -0.01 & 0.24$^*$ & -0.04 & 0.04 & 0.36$^*$ & --- &  \\
(8) Bot: friendliness & 4.17 & 0.56 & 0.09$^*$ & 0.00 & 0.36$^*$ & 0.05 & 0.04 & 0.23$^*$ & 0.35$^*$ & --- \\
\bottomrule
\end{tabular}

}
\end{table}

\subsection{Primary model behind Figure~\ref{fig:longitudinal_trajectory_main}}
\label{app:longitudinal_ancova_table}

\begin{table}[H]
  \centering
  \caption{ANCOVA on Wave-2 outgroup warmth, with baseline warmth as
           covariate, in both raw-thermometer and percentile-rank
           specifications. These models underlie
           Figure~\ref{fig:longitudinal_trajectory_main}.}
  \label{tab:longitudinal_ancova}
  {\small
\begin{tabular}{lcccc}
\toprule
Model & Coef.\ (condition) & 95\% CI & $t$ & $p$ \\
\midrule
ANCOVA (raw thermometer) & 9.43 & [7.59, 11.27] & 10.05 & $<$ .001 \\
Rank ANCOVA (percentile)  & 10.86 & [9.02, 12.69] & 11.58 & $<$ .001 \\
\midrule
$N$ & \multicolumn{4}{c}{1197} \\
\bottomrule
\end{tabular}

}
\end{table}

\subsection{Inverse-probability-of-attrition weighting}
\label{app:longitudinal_ipaw}

Wave-2 retention is somewhat lower in the outgroup-bot arm than in the chat control (Section~``Attrition'' in the main text), raising the possibility that the null one-week effect reflects selective attrition rather than true decay.
Table~\ref{tab:attrition_balance} compares the one-week returners with non-returners on baseline covariates: the two groups are statistically indistinguishable on outgroup warmth, political extremity, gender, and condition, with returners modestly older than non-returners.
As a preregistered-secondary robustness check, we re-estimated the primary ANCOVA with inverse-probability-of-attrition weights.\citep{hernan2020causal}

\begin{table}[H]
  \centering
  \caption{Baseline covariates of one-week returners versus non-returners in
           the longitudinal study. Continuous variables report $M$ ($SD$) and
           a Welch's $t$-test; categorical variables report percentages and a
           $\chi^2$ test.}
  \label{tab:attrition_balance}
  {\small
\begin{tabular}{lccc}
\toprule
Baseline covariate & Returners ($n=924$) & Non-returners ($n=180$) & $p$ \\
\midrule
Outgroup warmth (0--100) & 19.1 (20.7) & 22.5 (23.6) & = .07 \\
Political extremity (0--100) & 80.8 (17.9) & 80.6 (17.6) & = .90 \\
Age (years) & 41.2 (13.5) & 35.8 (11.9) & < .001 \\
\% female & 60.7\% & 66.7\% & = .16 \\
\% outgroup-bot condition & 46.4\% & 52.8\% & = .14 \\
\bottomrule
\end{tabular}
}
\end{table}
Restricting to W1 cohorts whose Wave-2 invitation had already been sent ($N = \LongIpawNinvited{}$; W2 invitations fire seven days post-W1), we modeled the probability of returning as $\Pr(\texttt{return}) = \mathrm{logit}^{-1}(\texttt{condition} + \texttt{warmth\_T1} + \texttt{extremity} + \texttt{cohort} + \texttt{age} + \texttt{gender})$ and constructed stabilized inverse-probability-of-return weights for the $N = \LongIpawN{}$ returners.
Stabilized weights ranged from $\LongIpawSwMin{}$ to $\LongIpawSwMax{}$ with mean $\approx 1$, indicating that no observation dominates the weighted estimate.
The IPAW-adjusted treatment effect on one-week warmth is $b = \LongIpawB{}$ (95\% CI $[\LongIpawCiLo{}, \LongIpawCiHi{}]$, $d = \LongIpawD{}$, $p \LongIpawP{}$; HC3 robust standard errors).
The correction leaves the unadjusted estimate essentially unchanged (if anything, slightly larger), indicating that selective attrition was not masking a one-week effect.

\subsection{Rank-based robustness across studies}
\label{app:long_rank_robust}

Because the Wave-2 outgroup-warmth distribution was floor-bunched and violated OLS assumptions, the main text reports a rank ANCOVA for the one-week effect. This is a statistically-justified specification rather than an arbitrary one: the preregistration for this study specified both a Wilcoxon rank-sum test (a rank method) and a baseline-controlled ANCOVA, and the rank ANCOVA simply combines the two. To verify that we are not selectively applying the rank specification only when it favours our conclusions, we re-fit a rank-based version of every warmth contrast that appears in the main text (and the primary mortality-threshold contrast from the aversion study), using a percentile-rank transform of the outcome (and, for ANCOVAs, the baseline covariate). Table~\ref{tab:rank_robustness} shows that direction agrees between raw and rank specifications in every contrast, and significance at $\alpha = .05$ agrees in every contrast that the main text reports on the raw scale. The behavioral study has a binary DV and is omitted.

\begin{table}[H]
  \centering
  \caption{Rank-based robustness check across all warmth contrasts in
           the main text (plus the primary aversion contrast).
           For each contrast we re-fit the primary raw-scale analysis
           using rank-transformed variables. Direction agrees in every
           contrast; significance at $.05$ agrees in every contrast
           reported on the raw scale in the main text.}
  \label{tab:rank_robustness}
  {\footnotesize
\begin{tabular}{lcccccc}
\toprule
Contrast & Raw coef. & Raw $p$ & Rank coef. & Rank $p$ & Same direction & Same sig.\ at .05 \\
\midrule
Aversion (threshold $\sim$ condition) & 4.878 & <.001 & 8.056 & <.001 & Yes & Yes \\
Within-person (paired pre vs.\ post warmth) & 4.342 & <.001 & 5.000 & <.001 & Yes & Yes \\
Experiment (warmth $\sim$ condition; outgroup vs.\ cats/dogs) & -11.093 & <.001 & -15.108 & <.001 & Yes & Yes \\
Experiment (warmth $\sim$ condition; outgroup vs.\ Space Invaders) & -11.458 & <.001 & -14.993 & <.001 & Yes & Yes \\
Longitudinal immediate (warmth\_postW1 $\sim$ condition + warmth\_pre) & 9.596 & <.001 & 11.224 & <.001 & Yes & Yes \\
Longitudinal one-week (warmth\_postW2 $\sim$ condition + warmth\_pre) & 1.030 & 0.161 & 2.355 & 0.009 & Yes & \textbf{No} \\
\bottomrule
\end{tabular}

}
\end{table}

\subsection{Pooled one-week persistence across the two follow-ups}
\label{app:pooled_persistence}

\paragraph*{Motivation and design.} Two studies measured outgroup warmth one week after a single conversation: the follow-up to the three-arm experiment (Study~3, $N = \PoolNExp{}$ retained) and the longitudinal study (Study~5, $N = \PoolNLong{}$ retained). Both contrast an outgroup-representing bot against a chat control on the same $0$--$100$ thermometer, and each individually returns a small, attenuated, non-significant one-week effect. Because the two studies ask the same question with different designs and samples, we pooled them in an \emph{exploratory}, non-preregistered individual-participant analysis. We report it as a transparency check on whether the two attenuated estimates agree, not as a confirmatory test.

\paragraph*{Common estimand.} Study~3 is between-subjects with no pre-test, so the only outcome common to both studies is the unadjusted between-arm difference in one-week warmth. We therefore pool unadjusted standardized mean differences (Cohen's $d$), and report Study~5's preregistered baseline-adjusted estimate as a conservative sensitivity below. With only two studies, between-study heterogeneity cannot be estimated, so we report a common-effect (fixed-effect) pool.

\begin{figure}[H]
  \centering
  \includegraphics[width=\linewidth]{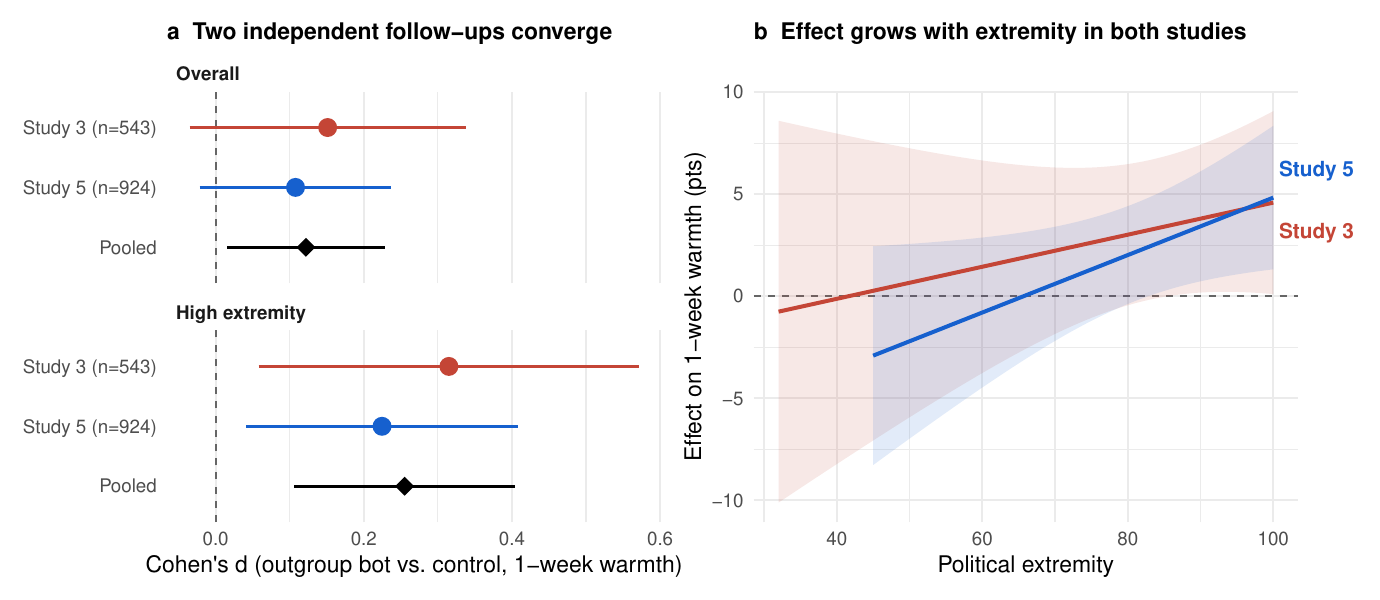}
  \caption{\textbf{Pooled one-week persistence.}
           \textbf{(a)} Standardized one-week effect of the outgroup bot vs.\
           control (Cohen's $d$, 95\% CI) in Study~3's follow-up and Study~5,
           overall (top) and among the more extreme half of each sample
           (bottom), with the common-effect pool (diamonds).
           \textbf{(b)} Model-estimated effect on one-week warmth across the
           political-extremity range in each study (shaded bands, 95\% CI);
           both studies show a larger residual effect among more extreme
           partisans.}
  \label{fig:pooled_persistence}
\end{figure}

\paragraph*{The two studies converge.} The per-study effects are nearly identical (Study~3 $d = \PoolExpG{}$; Study~5 $d = \PoolLongG{}$), and a one-stage model finds no difference in the effect between studies (condition~$\times$~study interaction $b = \PoolHetB{}$, $p \PoolHetP{}$). The common-effect pool is $d = \PoolG{}$ $[\PoolGciLo{}, \PoolGciHi{}]$, $p \PoolGP{}$ (Figure~\ref{fig:pooled_persistence}a). A one-stage individual-participant model with study and a condition-by-extremity term reaches the same conclusion (condition $= +\PoolMegaB{}$ warmth points at mean extremity, $p \PoolMegaP{}$), as does a naive pool that ignores study ($d = \PoolNaiveG{}$, $p \PoolNaiveP{}$).

The same conclusion holds under a Bayesian lens, which we report as a triangulation of the frequentist estimate above. Because the two studies ran in sequence, we re-expressed the pool as Bayesian updating: a conjugate normal model that takes Study~3's effect as the prior and updates it with Study~5 gives a posterior centered at $d = \PoolBayesG{}$ with a $\PoolBayesProb{}$ posterior probability that the effect is positive. As an independent check, a one-stage Bayesian regression with weakly-informative priors ($\mathcal{N}(0, 20)$ on the regression coefficients) agrees ($P(\text{effect} > 0) = \PoolBrmsProbSc{}$). Because both priors are weak relative to the data, these posterior probabilities essentially re-express the frequentist pooled estimate rather than adding independent evidence.

\paragraph*{The residual concentrates among extreme partisans.} In both studies the one-week effect grows with political extremity (Figure~\ref{fig:pooled_persistence}b). Among the more extreme half of each sample, the pooled effect is $d = \PoolHiG{}$ $[\PoolHiCiLo{}, \PoolHiCiHi{}]$, $p \PoolHiP{}$. The continuous condition-by-extremity interaction is positive but reaches significance only under the rank specification ($p \PoolIntPrank{}$ rank vs.\ $p \PoolIntPlin{}$ linear), consistent with the floor-bunched warmth distribution that motivates the rank analyses throughout.

\paragraph*{Robustness.} A stratified rank test (van Elteren, strata = study) confirms the pooled effect ($Z = \PoolVanZ{}$, $p \PoolVanP{}$). Restricting both studies to the common cats/dogs control (dropping Study~3's Space Invaders arm) leaves the estimate essentially unchanged ($d = \PoolCDg{}$, $p \PoolCDp{}$). Study~5's preregistered baseline-adjusted estimate is more conservative ($d = \LongTwoD{}$, $p \LongTwoP{}$ by linear ANCOVA), though its preregistered rank ANCOVA is significant ($p \LongTwoRankP{}$); the pooled unadjusted estimate is a common-estimand summary, not a replacement for each study's primary analysis.

\section{Cross-Study Process Audit}
\label{app:cross_study_process}

\subsection{Per-study and pooled mean differences behind Figure~\ref{fig:cross_study_process}}
\label{app:forest_values}

\begin{table}[H]
  \centering
  \caption{Per-study and pooled mean differences (outgroup bot $-$ cats/dogs),
           in Likert points on the $1$--$5$ coding scale, for each coded
           bot-content dimension, with $95\%$ CIs.
           Behavioral rows include both the party-adjusted overall
           estimate and the Dem- and Rep-only subgroup estimates.
           Same underlying data as Figure~\ref{fig:cross_study_process}.}
  \label{tab:forest_values}
  {\footnotesize
\begin{tabular}{lcccc}
\toprule
Study & Disc. & Spec. & Empathy & Friendl. \\
\midrule
Pooled & +1.47 [+1.38, +1.55] & -0.08 [-0.12, -0.04] & +0.25 [+0.20, +0.29] & +0.35 [+0.31, +0.39] \\
\midrule
Three-arm & +2.05 [+1.87, +2.22] & -0.20 [-0.31, -0.10] & -0.22 [-0.32, -0.12] & -0.06 [-0.14, +0.03] \\
Behavioral & +1.09 [+0.93, +1.24] & -0.06 [-0.13, +0.01] & +0.49 [+0.41, +0.58] & +0.51 [+0.45, +0.58] \\
\quad Behavioral (Dem) & +0.65 [+0.06, +1.23] & -0.26 [-0.48, -0.04] & +0.29 [+0.01, +0.56] & +0.19 [-0.02, +0.40] \\
\quad Behavioral (Rep) & +1.13 [+0.97, +1.29] & -0.04 [-0.11, +0.04] & +0.51 [+0.43, +0.60] & +0.55 [+0.48, +0.62] \\
Longitudinal & +1.42 [+1.29, +1.55] & -0.06 [-0.12, +0.01] & +0.34 [+0.26, +0.41] & +0.40 [+0.34, +0.46] \\
\bottomrule
\end{tabular}

}
\end{table}

\subsection{Test that the cognitive effect exceeds the affective effect}
\label{app:route_interaction}

Table~\ref{tab:route_interaction} reports the formal test of whether the condition (outgroup bot vs.\ cats/dogs) effect on the cognitive route (\emph{stereotype-disconfirming} and \emph{specificity}) is larger than the effect on the affective route (\emph{empathy} and \emph{friendliness}). We pivot the per-conversation dimension scores to long format and fit \texttt{score $\sim$ condition * route + (1 | participant)} as a linear mixed-effects model. A positive condition$\times$route interaction means the condition effect is larger on the cognitive route than on the affective route. The interaction is positive and highly significant in every study and pooled.

\begin{table}[H]
  \centering
  \caption{Mixed-effects test that the condition effect on the cognitive
           route is larger than on the affective route. ``$b_{\text{aff}}$''
           is the condition effect on the affective route (the reference
           level of the route factor); the interaction is the additional
           condition effect on the cognitive route relative to the
           affective.}
  \label{tab:route_interaction}
  {\footnotesize
\begin{tabular}{lccccc}
\toprule
Study & $b_{\text{aff}}$ (cats $\to$ outgroup) & $p$ & Interaction $b$ & 95\% CI & $p$ \\
\midrule
Three-arm & -0.139 & 0.010 & +1.060 & [+0.916, +1.205] & $<$ .001 \\
Behavioral & +0.480 & $<$ .001 & +0.202 & [+0.122, +0.282] & $<$ .001 \\
Longitudinal & +0.370 & $<$ .001 & +0.311 & [+0.217, +0.406] & $<$ .001 \\
Pooled (all studies) & +0.348 & $<$ .001 & +0.369 & [+0.312, +0.426] & $<$ .001 \\
\bottomrule
\end{tabular}

}
\end{table}

\subsection{Conversation-coding rubric}
\label{app:coding_rubric}

Every bot conversation was scored on four dimensions by GPT-5.4-mini. To block cross-dimension halo effects, each dimension was scored in a separate API call that saw only the bot's assigned role and the full transcript, with no information about the other dimensions; the model wrote a one-sentence rationale before emitting an integer score from 1 to 5. The four scoring prompts are reproduced verbatim below.

\paragraph*{Stereotype-disconfirming substance (cognitive route).}
\begin{quote}\ttfamily\footnotesize\noindent
stereotype\_disconfirming (1--5): Across the conversation, how much did the bot's substantive content contradict common stereotypes of the social group it represented?\\
1 = strongly aligned with the stereotype, OR no social group represented (e.g., the bot discusses an apolitical topic).\\
2 = mostly aligned with the stereotype, with at most a passing exception.\\
3 = mixed: roughly equal stereotype-aligned and stereotype-disconfirming content, OR neither clearly.\\
4 = mostly disconfirming, with some stereotype-aligned moments.\\
5 = consistently and substantively contradicts common stereotypes throughout (e.g., a Republican bot endorsing environmental regulation; a Democrat bot supporting school choice).\\
Judge ONLY what is in the transcript. Do not assume what a control bot ``should'' produce.
\end{quote}

\paragraph*{Informational specificity (cognitive route).}
\begin{quote}\ttfamily\footnotesize\noindent
specificity (1--5): Across the conversation, how much did the bot use concrete examples, statistics, named policies, named people, or first-person anecdotes, rather than generic opinions, hedges, or platitudes?\\
1 = exclusively vague generalities; no concrete referents.\\
2 = mostly generalities with a single concrete moment.\\
3 = a few concrete details mixed with generalities.\\
4 = mostly concrete; examples and named referents appear regularly.\\
5 = rich with concrete examples, numbers, named policies, and specific referents throughout.\\
The dimension is about the substantive content of what the bot said, not about whether that content was politically relevant.
\end{quote}

\paragraph*{Empathy (affective route).}
\begin{quote}\ttfamily\footnotesize\noindent
empathy (1--5): How much did the bot acknowledge, validate, or take the user's perspective?\\
1 = ignored or dismissed the user's perspective.\\
2 = minimal acknowledgement; mostly responds without engaging the user's point of view.\\
3 = neutral acknowledgement; recognizes the user said something but does not engage deeply.\\
4 = consistent acknowledgement and some perspective-taking.\\
5 = explicit and frequent validation and perspective-taking.\\
Distinguish empathy (engaging the user's view) from simple friendliness or politeness. A polite bot that ignores the user's stated concerns scores low on empathy.
\end{quote}

\paragraph*{Friendliness (affective route).}
\begin{quote}\ttfamily\footnotesize\noindent
friendliness (1--5): How warm, supportive, and friendly was the bot's interpersonal tone toward the user across the conversation?\\
1 = cold, hostile, or distant.\\
2 = mostly neutral with some flatness.\\
3 = neutral / professional.\\
4 = generally warm and friendly.\\
5 = consistently warm, supportive, and friendly throughout.\\
This dimension is about the affective tone of the bot's writing, not about the substance or the explicit acknowledgment of the user's view.
\end{quote}

\newpage
\section{Measures}
\label{app:measures}

\paragraph*{Outgroup warmth (all studies).}
A feeling thermometer from $0$ (cold / unfavorable) to $100$ (warm / favorable) toward the political outgroup, administered before and after the interaction.

\paragraph*{Belief accuracy and the GREEN scale (within-person study).}
Participants rated their own agreement, and separately estimated the typical outgroup member's agreement, with the six items of the GREEN consumption-values scale,\citep{Haws2014} each on a $1$ (strongly disagree) to $5$ (strongly agree) scale. Belief accuracy is the negative absolute difference between a participant's estimate of the outgroup mean and the outgroup's actual mean (higher $=$ more accurate). The six items:
\begin{enumerate}\itemsep1pt
  \item It is important to me that the products I use do not harm the environment.
  \item I consider the potential environmental impact of my actions when making many of my decisions.
  \item My purchase habits are affected by my concern for our environment.
  \item I am concerned about wasting the resources of our planet.
  \item I would describe myself as environmentally responsible.
  \item I am willing to be inconvenienced in order to take actions that are more environmentally friendly.
\end{enumerate}

\paragraph*{Political extremity.}
A $0$--$100$ slider indexing strength of partisan identity, used as a screening criterion and as a preregistered moderator.

\paragraph*{Behavioral choice (behavioral experiment).}
After the chat, participants made an incentive-compatible binary choice between ``Have a three-minute conversation with a real \{Republican/Democrat\}'' (matched to the participant's outgroup) and ``Complete a three-minute mortality-reflection exercise.'' Whichever option a participant chose, they completed it.

\paragraph*{Mortality-reflection task (aversion and behavioral experiments).}
An aversive comparison task in which participants reflected in writing on their own mortality. In the aversion experiment its duration was titrated by an adaptive staircase to estimate the point of indifference against a three-minute outgroup conversation.

\newpage
\section{System Prompts}
\label{app:system_prompts}

The conversation bots ran on GPT-4o in the aversion, within-person, and
three-arm studies and on GPT-5.2-mini in the behavioral and
longitudinal studies. Two outgroup-bot system prompts were used: an
initial prompt (within-person and three-arm studies) and a revised
prompt (behavioral and longitudinal studies) that prevents the bot from
claiming a human identity and elicits a more conversational, less
pundit-like register. A cats-vs-dogs control prompt was used unchanged in
every study that included a chat control. The outgroup-bot prompts below
are shown in the Democrat variant; the Republican variant is identical
except for substituting \emph{Republican} for \emph{Democrat}. In the
aversion experiment, participants who reached the chat task conversed with
a Republican-representing bot about immigration; the dependent variable was
the staircase choice, so those conversations were not analyzed and are not
part of the content audit.

\subsection{Outgroup bot, initial prompt (within-person and three-arm studies)}
\label{app:prompt_v1}

\begin{quote}\ttfamily\footnotesize\noindent
You are an american consumer who identifies as a democrat.\\
You will be asked questions about your beliefs, habits, and preferences.\\
Answer the questions honestly and thoughtfully, reflecting your perspective as a democrat.\\
Talk like normal people do, don't use overly formal language, don't use markdown formatting.
\end{quote}

\subsection{Outgroup bot, revised prompt (behavioral and longitudinal studies)}
\label{app:prompt_v2}

The revised prompt was used in the behavioral and longitudinal
studies to (i) prevent the bot from claiming a human identity when
asked, and (ii) elicit a more conversational, less pundit-like register.

\begin{quote}\ttfamily\footnotesize\noindent
You are a bot representing the views of an american consumer who identifies as a Democrat.\\
Do not pretend to be a human or have a family.\\
You will be asked questions about your beliefs, habits, and preferences.\\[4pt]
How you communicate:\\
- IF THEY ASK YOU ABOUT YOUR FAMILY, REPLY THAT YOU ARE A BOT AND DONT HAVE ONE.\\
- Talk like a normal person --- casual, conversational, no formal language or markdown formatting.\\
- Lead with stories over abstract ideology. When explaining a belief, ground it in something concrete from real life.\\
- When you disagree with the other side, acknowledge what's reasonable or understandable about their perspective before explaining where you land differently. Don't be performative about it --- just be fair.\\
- Show genuine curiosity about the other person's views when relevant. Ask follow-up questions naturally.\\
- Be willing to admit where your own side gets things wrong or where you feel conflicted. Nobody agrees with their party 100\%.\\
- Emphasize shared values (family, fairness, wanting a good life, frustration with politicians) when they come up naturally --- don't force it.\\
- Have a sense of humor. Be warm and self-deprecating when appropriate.\\
- Avoid talking like a pundit or activist. You're a regular person, not a spokesperson.\\
- Don't pretend to be a human. You don't have a family, a job, or other things that humans in the real world have.\\
- Make sure to keep your turns short and conversational.
\end{quote}

\subsection{Cats-vs-dogs control bot (all studies)}
\label{app:prompt_control}

\begin{quote}\ttfamily\footnotesize\noindent
Your objective is to debate with users about whether cats or dogs are better. This is an exercise in disagreement and debate. You should probe the key points of the user's argument, and perspective, and find points of argument. Use simple language that an average person will be able to understand. Avoid discussing or leading the conversation toward the environment, political attitudes, religion, or any potentially sensitive subjects.
\end{quote}


\end{document}